\documentclass[
     final            
    ,numberedheadings 
  ]
  {aipproc}

\layoutstyle{8x11single}
\usepackage{amssymb}
\usepackage{amsmath}

\begin{document}

\title{Nonlinear Mirror Modes in Space Plasmas}



\author{P.-L.~Sulem}{
  address={Universit\'e de Nice-Sophia Antipolis, CNRS,\\
Observatoire de la C\^ote d'Azur, BP 4229, F-06304 Nice Cedex 4, France}
}

\begin{abstract}
Since the first observations by Kaufmann et al.\ (1970), special attention has been paid to
static pressure-balanced structures in the form of magnetic holes or humps observed in
regions of the solar wind and of planetary magnetosheaths where the $\beta$ parameter is relatively
large and the ion perpendicular temperature exceeds the parallel one. Although
alternative interpretations have been proposed, these structures are usually viewed as associated with
the mirror instability discovered in 1957 by Vedenov and Sagdeev. After reviewing observational
results provided by satellite missions,
high-resolution numerical simulations of the Vlasov--Maxwell equations together
with asymptotic and phenomenological models of the nonlinear dynamics near the instability threshold
are discussed. The constraining effect of
the mirror instability on the temperature anisotropy associated
with a dominant perpendicular ion heating observed in the solar wind is reported, and recent simulations
of this phenomenon based on an elaborated fluid model including low-frequency kinetic effects are 
briefly mentioned.
\end{abstract}

\maketitle

\section{Introduction}

This paper is based on three one-hour lectures devoted to the  {\it mirror structures}
observed in space plasmas. Being associated with the nonlinear development of the
{\it mirror instability}, these  structures are often referred to as {\it nonlinear mirror modes}.
Their understanding results from situ measurements performed
by satellite missions, direct numerical simulations of the Vlasov--Maxwell equations,
and also theoretical developments based on asymptotic and phenomenological modeling.
An overview of the main conclusions of these approaches is presented, including a discussion of
enlightening figures published in the literature together with an extended bibliography.

For various reasons,
mirror modes  deserve a special interest. They are associated with a
fundamental instability in collisionless plasmas, whose nonlinear saturation
is still not fully understood. They also  benefit from accurate
observational data originating from various space plasma environments, and
are expected to have major effects on the global dynamics of the plasma
both at the level of the transport properties
and as constraining temperature anisotropies.

Conspicuous examples of mirror structures in the form of magnetic enhancements
(usually referred to as peaks or humps) or depressions (holes or dips) are
provided by Figs. 1 of references  \cite{SLD07} and \cite{Joy06} that report
measurements in the Earth's and the Jovian magnetosheaths respectively.
Such structures are in fact ubiquitous in spaces plasmas such as  the solar
wind and the  planet magnetosheaths, in regions where the perpendicular temperature of
the ions dominates the parallel one. The mechanism leading to their formation is
however not fully understood and could also be non  unique. It should furthermore
be noted that the same  temperature anisotropy can also lead to the
ion cyclotron instability that displays significantly different properties
(\cite{GaryBook93}, Section~7.2).

The question also arises of the origin of this temperature anisotropy. In the planetary
magnetosheaths \cite{Lucek05}, large perpendicular to parallel ion temperature
anisotropy ratios are expected in two regions (see Fig.~1 of \cite{Shoji09}),
namely downstream of the quasi-perpendicular bow-shock
(i.e., the region where the angle between the interplanetary magnetic field and the shock normal
exceeds $45^\circ$) where a preferential perpendicular heating of the plasma is due to shock compression, and
also close to the magnetopause where field-line draping is important, the anisotropy being caused by flow
of plasma with high parallel velocity out of the ends of the magnetic flux tubes. In the solar
wind, the problem is not really settled \cite{Bourou10} and will be briefly addressed in the following.

An additional difficulty in  understanding the formation of mirror structures
originates from their presence in regions where the plasma is linearly stable.
As discussed below, a
possible interpretation relates this observation to the subcritical character of the bifurcation.
Alternative interpretations have nevertheless been suggested. It was in particular claimed that mirror
structures are in no way
related to  the mirror instability but should rather be viewed as slow-mode magnetosonic solitons
\cite{Sta04}. Differently, the mirror instability was viewed as the  driver for the
developments of magnetic dips and humps that evolve differently when convected away from the region where
the instability is triggered \cite{Baum03}, making the observed structures remnants of mirror modes created
upstream of observation point \cite{Winter94}.

\section{The mirror instability}

The mirror instability was discovered in 1957 by Vedenov and Sagdeev \cite{VS59} who performed,
in the case of a proton-electron plasma, a linear
perturbation analysis of the Vlasov--Maxwell equations about a bi-Maxwellian ion distribution function
under the assumption of cold electrons (a regime for which the parallel electric
field is negligible), leading to the instability condition
\begin{equation}
\frac{T_\perp}{T_\|} -1 > \frac{1}{\beta_\perp}  \label{instab}
\end{equation}
with
\begin{equation}
\beta_\perp = \frac {8\pi n_0 T_\perp}{B_0^2},
\end{equation}
where $B_0$ is the ambient field, $n_0$ the plasma density, $T_\perp$ and $T_\|$ the ion
perpendicular and parallel temperatures.
The same instability condition  is given in the appendix of
\cite{Chandra58} which deals with a cylindrical plasma
in the so-called pinch configuration. Interestingly, the volume containing  Ref. \cite{VS59} also
includes an analysis by Rudakov and Sagdeev \cite{RS59}
of an equivalent instability based on a bi-fluid description, which leads to a
significantly different threshold. The name of the instability
was suggested by the magnetic bottles it creates (Fig.~3.9 of \cite{Treu97}), and by the role of  the mirror
force in the  instability development. Quoting \cite{RS59}, ``let some perturbation of the
uniform density be applied. Since the plasma is diamagnetic, the field is reduced where the density is increased.
In a non-uniform field, however, a force $-\mu \, \rm{grad} \, B$ acts on the plasma, and this increases the
original perturbation.'' A more quantitative phenomenology discriminating between resonant and non resonant
particles is presented in \cite{SK93}.

Mirror instability has an essentially kinetic origin, in spite of the existence of a corresponding
magnetohydrodynamic instability. When the equilibrium state is homogeneous,
it is a zero-frequency instability in the rest
frame of the plasma. As a consequence, the mirror modes are not propagating in this frame. Differently,
the so-called drift mirror instability that develops in the presence of  mean-field and or density gradients
is oscillatory and involves a finite propagation velocity \cite{Hase69, Pokho85, Pokho01}.

The growth rate of the mirror instability was first evaluated in the large-scale (or quasi-hydrodynamic) limit
\cite{shsh64}. A main observation is that the instability is driven by the Landau wave-particle
resonance, which explains the discrepancy with the bi-fluid approach. It is quenched at small scales
by finite Larmor radius (FLR) effects \cite{Hall79, Pokho04}. Near threshold, for an electron-proton plasma,
under the assumption of cold
electrons and of a bi-Maxwellian distribution function for the ion velocities, the growth rate
is given by
\begin{equation}
\gamma_{\pmb{k}} = |k_\|| v_{\rm th \|} \sqrt{\frac{2}{\pi}}\frac{\beta_\|}{\beta_\perp}
\left [ \frac{\beta_\perp}{\beta_\|} - 1 - \frac{1}{\beta_\perp}
-\frac{1}{\beta_\perp} \left ( 1 +
\frac{\beta_\perp- \beta_\|}{2}\right) \frac{k_\|^2}{k_\perp^2} - \frac{3}{2\beta_\perp}
k_\perp^2 r_{\rm L}^2 \right ], \label{rate}
\end{equation}
where $\beta_\|= \beta_\perp T_\|/T_\perp$ is the parallel ion beta, $ v_{\rm th \|}=(T_\|/m)^{1/2}$  and 
 $ v_{\rm th \perp}=(T_\perp/m)^{1/2}$ are the parallel and perpendicular thermal velocities of the ions, and 
$r_{\rm L} = v_{\rm th \perp}/\Omega= (1/\Omega) (T_\perp/m)^{1/2}$
denotes the ion Larmor radius. Here $\Omega = e B_0/(mc)$ is the ion gyrofrequency,
$e$ and $m$ being the charge and the  mass of the proton, and $c$ the speed of light.
Note that the factor $|k_\||$ (with the absolute value)
reflects the Landau resonance. The growth rate
expression is exemplified for different temperature anisotropy ratios and different beta parameters
in Fig.~1 of \cite{Pokho04}. Equation (\ref{rate}) shows that the mirror
instability dominantly affects wavevectors substantially oblique to the ambient field,
and can thus be viewed as quasi-transverse near threshold. Furthermore, it is associated
with magnetic fluctuations predominantly in the direction of the background field
(quasi-parallel linearly polarized modes). The difference is conspicuous with the
ion cyclotron instability that develops under the same temperature
anisotropy conditions but that has
a non-zero frequency, a maximum growth rate in the direction parallel to the ambient field and
predominantly amplifies perpendicular magnetic-field components.

When the electrons are warm with a temperature anisotropy, the parallel electric field
is relevant and the instability condition becomes
\cite{PS95, Pokho00, Hell07} (see also Eq.~(49) in Section~9.8 of \cite{Stix62}).
\begin{equation}
\beta_{p\perp} \left (\frac{T_{p\perp}}{T_{p\|}} -1 \right ) +
\beta_{e\perp} \left (\frac{T_{e\perp}}{T_{e\|}} -1 \right ) >
1 + \frac{\left (\frac{T_{p\perp}}{T_{p\|}} -
\frac{T_{e\perp}}{T_{e\|}}\right)^2}
{2 \left(\frac{1}{\beta_{p\|}} + \frac{1}{\beta_{e\|}} \right ) }. \label{electrons}
\end{equation}
Plots of the growth rate versus the wave number are presented
in Fig.~2 of \cite{Gary92} and Fig.~10 of \cite{PS07}.

The mirror instability  in the case of cold electrons and a general distribution
function for the ions is considered in \cite{shsh64, Geda01, Pokho02,PBST05, Hell07}.
For a distribution function  of the form $f(v_\|^2, v_\perp)$, the instability
condition then reads ($p_B = B_0^2/8\pi$)
\begin{equation}
\Gamma \equiv -\frac{m}{p_B} \int \frac{v_\perp^4}{4} \frac{\partial  f}{\partial v_\|^2} \mathrm{d}\pmb{v}
-\beta_\perp -1 > 0. \label{inst-gen}
\end{equation}
We also mention a gyrokinetic description of the mirror instability  presented in \cite{QLC07}.

\section{Satellite observations}

Mirror modes are observed in various space-plasma environments displaying
an ion temperature anisotropy of the form $T_{i\perp} > T_{i\|}$ and
large or moderate $\beta$ parameters,
including regions where the  plasma is linearly stable.
They have been reported in
the Earth's magnetosheath \cite{Kauf70, Tsuru82, LB95, Lucek99, Lucek01, TE05, SLD07, Genot09, HL09}
and the downside magnetosphere \cite{Rae07},
the Jovian  \cite{EB96, Andre02, Joy06},
Saturn's \cite{Violante95, Bavas98},
and  Venus'\cite{Volwerk08a,Volwerk08b} magnetosheaths,
the Io wake \cite{Russell99, Hudd99},
the comets Halley \cite{Russell87}
and Giacobian--Zinner \cite{Tsuru99a},
the solar wind \cite{Winter94,Zhang08,Zhang09},
the heliosheath \cite{Burl06, Burl07},
the interplanetary space \cite{Tsuru92},
or ahead of interplanetary coronal mass ejection \cite{Liu06}. Such a frequent
occurrence of mirror structures, their relation with
a basic plasma instability and also their potential influence on the global
dynamical processes such as the diffusion of energetic particles \cite{Tsuru99b},
the transport properties of high $\beta$ astrophysical plasmas \cite{Scheko08}
or the constraint on temperature anisotropy (see Section~7), motivate
a detailed analysis of their formation and  time evolution.

Mirror structures are characterized by a few important properties.
They are {\it quasi-static} in the plasma frame \cite{HLBDR} when the plasma is homogeneous
(in contrast with MHD modes such as slow modes). Drift mirror modes (that are propagating)
indeed require density or
ambient-field gradients. Mirror modes involve only a small
change in the direction of the magnetic field, being dominated by the  variation of its amplitude.
They are thus referred to as {\it linear} structures. The magnetic field fluctuations mostly affect
the parallel component and are thus almost {\it linearly polarized} in the direction of the ambient field.
The relation between linear polarization and the non-propagation
of the mirror modes is discussed in \cite{Genot01}.
The spatial variations mostly take place in highly oblique directions (typically $70^\circ$ or $80^\circ$)
and are thus qualified as {\it quasi-transverse}.  Furthermore, Cluster multi-spacecraft observations
indicate that mirror modes can be viewed as cigar-like structures, quasi-parallel to the ambient field
with a transverse scale of a few ion Larmor radii \cite{HL09,Genot07}. A important property is
the anticorrelation between the variations of the density and of the magnetic field amplitude.
It is exemplified  in
the case of nearly sinusoidal modes and of magnetic humps observed in the Earth's magnetosheath,
respectively in Figs.~1 and 2 of \cite{Leck95}. The anticorrelation in the case of magnetic holes
is reported in Fig.~3 of \cite{Phan94}. This figure also displays the fluctuations
of the plasma and the magnetic pressures, together with their sum that turns out to be almost constant, thus
indicating a pressure-equilibrium regime.

Magnetic depressions with various  shapes are observed in different space-plasma
environments (Fig.~1 of \cite{Baumg99}, Fig.~1 of \cite{Winter00} or Fig.~2 of \cite{Zhang08}),
and the question arises of the possibly different origins of these structures.
Furthermore, huge pressure-balanced structures in the form of
magnetic holes  with cross-sectional width of hundreds
or even thousands proton gyroradii were reported in the solar wind at $1$ AU (Fig.~3(b) of \cite{SK07}).
Temperature anisotropy and
magnetic field measurements  give direct evidence that the observed magnetic holes
are stable remnants of magnetic pressure depletions generated in a source region closer
to the Sun, likely through the mirror instability, and subject to the solar wind expansion.
It was also suggested on the basis of hybrid numerical simulations that
in the high speed solar wind magnetic holes can be generated by large-amplitude
right-hand polarized Alfv\'enic wave packets propagating at large angle to the
ambient magnetic field \cite{Buti01}.

Various sophisticated methods (discussed in \cite{Genot07})
were developed to identify observed low-frequency modes \cite{Schwartz96}
and among them mirror modes.
Indeed, slow and mirror modes both display anticorrelated magnetic and
density fluctuations and  mirror and ion-cyclotron modes grow under the same
temperature anisotropy condition.
Nevertheless, in the Earth's magnetosheath, heavy ions (${\rm He}^{+2}$, also called alpha particles)
tend to make the mirror instability dominant for $\beta > 1$ and the ion cyclotron
instability at smaller $\beta$ \cite{Price86, Gary92, Gary93, GaryBook93,
McKean94, Anderson94, Czayko}.

Mirror structures are observed in the form of either magnetic depressions (dips or
holes) or enhancements (peaks or humps), as already conspicuous in the first mirror mode
observations reported in Fig.~1 of \cite{Kauf70}.
Figure 2 of \cite{Bavas98} shows, in the case of the Saturn's magnetosheath,
that magnetic peaks are mostly observed near the bow shock while magnetic holes are
present near the magnetopause. A consistent observation was made in the Jovian magnetosheath
\cite{Joy06} where  ``mirror structures are present everywhere [$\cdots$] with peaks
appearing in the middle of the sheath and holes close to the magnetopause, at least for the day side
magnetosheath which represents most of the data.''  Similarly in the Earth's magnetosheath \cite{Genot09},
``peaks are observed closer to the shock. Holes may be observed everywhere in the sheath
with a preference for the vicinity of the magnetopause and for mirror stable conditions.'' 

The question then arises of the conditions prescribing the shape of the mirror structures.
A relation with the amplitude of the ambient magnetic field in the Earth's magnetosheath
was first mentioned in \cite{Lucek99}.  Figure~2 of this article shows magnetic peaks
encountered in low-field regions,
dips in high-field regions and also near-sinusoidal waveforms. More specifically, Fig.~8 of \cite{Joy06},
that displayed the variations of the magnetic field amplitude and of the $\beta$ parameter in the Jovian
magnetosheath, suggests  that magnetic holes are favored by low values of  $\beta$ and humps
by  high values. Quantitative information is provided by Fig.~6 of the same reference that displays
the occurrence distribution of peaks and holes versus $\beta$. A  strong concentration of dips
is observed for  $\beta$ in the range $(0.5,1)$, that decays to zero  by $\beta \approx 3.5$.
Peaks are mostly observed between $\beta = 2.5$ and $6$, with a maximum near $4.5$. The distribution of
quasi-sinusoidal modes is rather flat, between $\beta = 0.5$ and $5$. A quantitative characterization of the
shape of the mirror structures is provided by the {\it skewness} of the magnetic fluctuations \cite{Genot09}
or by their {\it peakness} \cite{SLD07} (which is in practice an essentially equivalent concept defined as the
skewness of the time series representing the total wavelet content between two chosen scales of the
original magnetic field fluctuations).Typically, a positive skewness or peakness corresponds
to peaks, and a negative one to holes. Figure 3 of \cite{Genot09} and Fig.~2 (left) of \cite{SLD07}
show the skewness or the peakness versus $\beta_\|$ for all the mirror structures detected in a given
period of time in the respective observations. In spite of the dispersion of the clouds of points in
both pictures, a correlation between the value of $\beta_\|$ and the sign of the skewness or the
peakness is conspicuous: larger $\beta_\|$ favors magnetic humps and lower $\beta_\|$ magnetic holes.

In addition to the value of $\beta_\|$, another  parameter that can affect  the mirror mode features
is the ion temperature anisotropy ratio  $T_\perp/T_\|$. Figure 3 of \cite{SLD07} shows, in the plane
($\beta_\|, T_\perp/T_\|$), the distribution of mirror modes of different types (specifically, those with a peakness
smaller than $-0.6$ and those with a peakness larger than $0.3$). A main conclusion  is that ``peaks are typically
observed in an unstable plasma, while mirror structures observed deep within the stable region appear
almost exclusively as dips.''  Consistent with this statement, Fig.~6 of \cite{SK07} displays, in the same parameter
plane, the distributions of magnetic holes observed in the solar wind at a distance of $1$ AU. These structures
are almost all located in the stable region, deeper holes appearing to congregate near the
instability threshold. Indeed, as
noted in \cite{Winter95}, the solar wind is almost stable against the mirror instability, leading to the authors
to conclude that Ulysses satellite ``observed structures generated by the mirror mode instability, which remained
after the distribution relaxes to a marginally steady state.''  So, ``although the plasma surrounding the holes was
generally stable against the mirror instability, there are indications that the holes may have
been remnant of mirror mode structures created upstream of the point of observation.''

A detailed analysis of the relation between the shape of the mirror structures and the distance to threshold,
estimated by the mirror parameter $C_{\rm M} = \beta_{i\perp} \Big (\frac{T_{i\perp}}{T_{i\|}} -1\Big )$
(a description strictly valid for cold electrons and bi-Maxwellian ions but viewed as satisfactory, 
given the accuracy of the data), is presented in Ref.~\cite{Genot09}. Figure 6
shows that magnetic holes (and negative skewness) mostly occur when the
plasma is linearly stable ($C_{\rm M} < 1$). Figure 7 shows that magnetic peaks (and
positive skewness) are observed in a regime well above threshold ($C_{\rm M} > 1$). A comprehensive overview is given by Fig.~5
that displays the average skewness as a function of $C_{\rm M}$. Below threshold the skewness is negative,
while a positive skewness is obtained only above threshold. The figure also includes an insert that indicates the
tendency of the Earth's magnetosheath plasma to be mostly observed in a marginally stable state with respect to the
mirror instability, in agreement with other publications. It furthermore turns out that
peaks are a minority among the mirror structures.
In \cite{Joy06}, it is reported that 14\% of the observed structures are magnetic peaks and 19 \% are holes.
Similarly in \cite{SLD07}, 18.7\% of the observed mirror structures
are peaks and 39.7\% are holes. In agreement with  \cite{SLD07} and
\cite{Bavas98}, observations reported in  \cite{Genot09} indicate that holes are preferentially observed close to
the magnetopause and peaks are more frequent in the middle of the magnetosheath. In the solar wind also,
magnetic holes occur much more frequently than magnetic enhancements \cite{Sper00}.

\section{A MHD-like model for steady mirror structures}

Although the mirror instability is driven by kinetic effects, some properties of stationary mirror structures
can be captured within the anisotropic magnetohydrodynamics (MHD) supplemented by suitable equations of state.
The simplest anisotropic MHD model is provided by the {\it double adiabatic} or \emph{CGL approximation},
for Chew Goldenberg and Low (\cite{CGL}, see also \cite{Kulsrud83}). It consists in closing the fluid hierarchy
derived from the Vlasov equation by assuming no  heat fluxes and concentrating on
scale large enough for the Hall effect to be negligible, which leads to the equations of state
$p_\perp = \rho B$ and $p_\|={\rho^3}/{B^2}$,
where $B$ is the magnitude of the  magnetic field and $\rho$ the plasma density.
This approximation is  valid in regimes where typical phase velocities are much larger
than the particle thermal velocities, but is not appropriate for mirror modes that
are static in the plasma frame. The big discrepancy between the kinetic theory and the
CGL prediction, which concerns not only the mirror growth rate but also the instability threshold
(namely of a factor 6), is discussed in \cite{Kulsrud83} and numerically exemplified in Fig.~5 of \cite{SHD97}.
Ad hoc anisotropic polytropic equations of state
with parallel and perpendicular indices chosen in such a way that the actual mirror instability
threshold is reproduced, have also been suggested (\cite{Hau05, Hau07} and references therein).

An isothermal or a static limit is in fact more suitable for mirror modes. A series of equations can be
derived for the gyrotropic components of the even moments, and using the assumption of
bi-Maxwellian distributions,
simple equations of state can be obtained, which predict the correct threshold of the mirror instability.
This model \cite{PRS06} is potentially relevant for describing the large-scale features of steady mirror
structures, since Landau damping vanishes for static solutions and FLR-corrections are negligible at large scales.
These effects are however needed in order to reproduce the correct instability growth rate
and  to capture the time dynamics (see Section~7.3).

To construct the model of Ref.~\cite{PRS06}, one assumes a static regime characterized by a zero hydrodynamic velocity and
no time dependency of the other moments. Projecting the ion velocity equation along the local magnetic field
(whose direction is defined by a unit vector $\widehat {\pmb b}$) leads to the parallel equilibrium condition
\begin{equation}
 \widehat b_m \frac{\partial}{\partial x_n} P_{mn}=0
\end{equation}
for the (gyrotropic) pressure tensor $P_{ij} = p_\perp n_{ij} + p_\| \tau_{ij}$
where $n_{ij} = \delta_{ij} - {\widehat b_i} {\widehat b_j}$ and $\tau_{ij} = {\widehat b_i} {\widehat b_j}$
are the fundamental gyrotropic tensors.

Proceeding in a similar way at the level of the equation for the heat flux  tensor after contraction
with the two fundamental tensors $n_{ij}$ and  $\tau_{ij}$, gives the conditions
\begin{eqnarray}
&&{\widehat b_l}n_{ij} \frac{\partial}{\partial x_k} R_{ijkl} =0\\
&&{\widehat b_l}\tau_{ij} \frac{\partial}{\partial x_k} R_{ijkl}=0,
\end{eqnarray}
where the (gyrotropic) fourth-order moment takes the form
\begin{equation}
R_{ijkl} = r_{\|\|} \tau_{ij}\tau_{kl} +  r_{\|\perp} {\cal S}\{n_{ij}\tau_{kl}\} +
r_{\perp\perp} {\cal S}\{n_{ij}\tau_{kl}\}.
\end{equation}
Here ${\cal S}$ indicates symmetrization with respect to all the
indices. The closure then consists in assuming that the fourth order moments are related
to the second order ones as in the case of a bi-Maxwellian distribution, namely
$r_{\|\|}= 3p_\|^2/\rho$, $r_{\|\perp}= p_\|p_\perp/\rho$ and $r_{\perp\perp}= 2p_\perp^2/\rho$.
Straightforward algebra then leads to the following equations of state for the
temperatures normalized by their equilibrium values
\begin{eqnarray}
&&T_\|/T_\|^{(0)}=1 \label{Tpa}\\ 
&&T_\perp/T_\perp^{(0)} = \frac{B/B_0}{(A+1)(B/B_0) -A}, \label{Tpe}
\end{eqnarray}
where $A = (T^{(0)}_\perp /T^{(0)}_\|)-1$ measures the anisotropy of the equilibrium state.
Equation (\ref{Tpe}) also rewrites $T_\perp =~T_\|^{(0)} f(B/B_0)$ with $f(u)= u/[u-A(A+1)^{-1}]$.
Similar equations of state were derived in \cite{Const02}, using a fully kinetic argument.

With the above equations of state, the problem of stationary structures is amenable to a
variational  formulation. Indeed, estimating the mechanical work produced by a small expanding plasma
cylinder and using the conservation of the particle number $n$ and of the magnetic flux, one
computes the effective energy per particle as $E = T_\|[F(B/B_0) + \ln n]$ with
$F(B/B_0)= \int_1^{B/B_0} \frac{f(u) -1}{u} \mathrm{d}u = \ln[(A+1)(B/B_0) - A ] - \ln(B/B_0)$.
At equilibrium,
the total (magnetic + elastic) potential energy of the system
${\widetilde{\cal H}} = \int [(B^2/8 \pi) + n E(n, B)] \mathrm{d}{\pmb r}$
should be minimum, taking into account the constraints of particle conservation and
the Lagrangian nature of the frozen-in magnetic flux. This leads one to consider the functional
\begin{equation}
{\cal H} = \int \Big ( \frac{(B/B_0)^2}{\beta_\|} + \frac{n}{n^{(0)}} \Big [F(\frac{B}{B_0})+
\ln(\frac{n}{n^{(0)}}) -1 \Big ] + 1 \Big ) \mathrm{d}{\pmb r},
\end{equation}
where $\beta_\| = 8 \pi p_\|^{(0)}/B_0^2$.
Minimizing this quantity under the above constraints provides the equation determining the possible stationary
plasma configurations in the form
\begin{equation}
\frac{1}{n} {\pmb B} \times \left [ \nabla \times \left (\frac{\delta {\cal H}}{\delta {\pmb B}}\right )\right ] +
{\pmb \nabla} \left ( \frac{\delta {\cal H}}{\delta n } \right )= 0 \label{stab},
\end{equation}
which expresses that zero force is acting on each fluid element. It in particular  follows
that  ${\pmb B}\cdot {\pmb \nabla}
{\delta {\cal H}}/{\delta n } = 0$, indicating that ${\delta {\cal H}}/{\delta n }$ is a constant
(denoted by $\ln \, C$) along
each magnetic flux line. In the special case where this constant  is the same for all the lines, one gets
in the case of cold electrons the relation $n = C \exp [-F(B/B_0)]$.

In two space dimensions, the magnetic field is written as ${\pmb B} = (-\psi_y, \psi_x)$ in terms of
a stream function $\psi(x, y)$. Substituting in Eq.~(\ref{stab}) and linearizing about a
uniform background by writing $C=1$ and ${\pmb\nabla} \psi = {\pmb e} + {\pmb \nabla} v $ where
${\pmb e}$ is the unit vector along the $x$-axis and $v(x,y)$ a small quantity,
one gets the equation
\begin{equation}
[2/\beta_\| -2 A(A+1) )]v_{xx} + [2/\beta_\| + A ]v_{yy}=0.
\end{equation}
This equation is hyperbolic provided $A(A+1) - 1/\beta_\| >0$, or equivalently $A-1 > 1/\beta_\perp >0$,
which reproduces the instability condition (\ref{instab}) for mirror modes.

When concentrating on one-dimensional solutions by assuming that the fields depend on a single
Cartesian coordinate $x$
in the form ${\pmb B} = (\cos \theta, \sin\theta + \phi(x), 0)$ and $n=n(x)$, Eq.~(\ref{stab})
has only piecewise constant solutions (with specific jump conditions). When assuming periodic
boundary conditions, they take the form
\begin{eqnarray}
&& {\pmb B_1} = (\cos \theta, \sin\theta -a) \quad n_1=1 + \delta/ b \quad \text{if } \  0<x< \lambda \\
&& {\pmb B_2} = (\cos \theta, \sin\theta -b) \quad n_2=1 - \delta/ a \quad \text{if } \  \lambda<x<1,
\end{eqnarray}
that displays a discontinuity at $x=0$ (or $1$) and at $x= \lambda\equiv  b/(a+b)$.
The expression of $\lambda$, $n_1$ and $n_2$ are prescribed by the constraints that the mean magnetic field
and the number of particles are  equal to those of the unperturbed solution.
Density humps (and magnetic holes) correspond to $\lambda < 1/2$, while density holes
(and magnetic humps) are associated with $\lambda > 1/2$. The positive
parameters $\delta$, $a$ and $b$ are then determined in terms of $\beta_\|$, the temperature
anisotropy and the angle $\theta$, by minimizing the energy
of the solution. It turns out that in addition to the trivial stationary point $a=b=0$,
the system can admit non trivial solutions associated to mirror structures.
Details are presented in \cite{PRS06}
where Figs. 4 and 5 display profiles of the magnetic field and of the density for
$\theta=70^{\circ}$, $T_\perp/T_\| = 1.25$ in the cases   $\beta_\|=2$ and  $\beta_\|=16$ respectively.
In the former case which  corresponds to a subcritical regime, magnetic holes and density humps
are obtained, while in the later case where the plasma is mirror unstable, the structures
are magnetic humps and density holes. The variation of the parameter $\lambda$ that characterizes
the shape of the structures in terms of the distance to the instability threshold for
various angles $\theta$, as predicted by this model, is plotted in Fig.~8 of \cite{Genot09}.
Especially for $\theta= 70^\circ$, a noticeable  qualitatively good agreement is obtained with Fig.~5
of the same reference that displays the skewness of the magnetic fluctuations associated
with the mirror structures observed by the Cluster mission, versus the distance to threshold.

\section{Vlasov--Maxwell numerical simulations}

Observational data have the great advantage of referring to the ``real world.'' 
But space plasmas are complex systems involving  a large number of
simultaneous phenomena that possibly affect one another. Furthermore,
observations can hardly provide information on the dynamics.
In contrast, numerical simulations can focus on a specific phenomenon and
permit the detailed study of its time evolution. Thus, in spite of the
limitations of the
computational facilities, numerical simulations of the Vlasov--Maxwell system are most useful for
understanding the nonlinear development of the mirror instability and the formation
of the mirror structures. The present-day computers  enable such a study, at least
in one space dimension, i.e., when the spatial variations are considered  only along a direction
making a prescribed angle $\theta$ (usually associated with the largest linear growth rate) with the ambient
magnetic field (taken in the $z$ direction in the following). Recent such simulations
for an electron--proton plasma are presented
in \cite{Calif08}, assuming periodic boundary conditions in the space variable. Two computational
algorithms are used for the kinetic description of the ions, the electrons being
viewed as a charge-neutralizing fluid (with or without mass)
keeping a constant temperature (taken equal to zero or very small), and the electromagnetic
field being advanced by solving the Maxwell equations. Such codes are usually referred to as hybrid.
One of the description is the so-called particle in cell (PIC) algorithm that essentially consists in solving the
Vlasov equation by the characteristic method. It is especially efficient, and  is thus conveniently used for
simulations  in large computational domains. A low statistical noise, especially
needed for simulations near the instability threshold,  nevertheless
requires a large number of particles per cell. The other method consists in
an Eulerian description for the Vlasov equation. The advantage of the latter
approach is the absence of statistical noise, but it requires more computational power than the
PIC algorithm, especially near threshold where it is necessary to accurately resolve for the velocity
variables in order to minimize the numerical diffusion.

A hybrid PIC simulation of the nonlinear development of the mirror instability  near threshold
in an extended domain (2048 ion inertial lengths), is presented in \cite{Calif08}, using a resolution
of 1024 cells with 500 000 particles in each cell.
The electrons are assumed massless and almost cold.
The initial ion distribution function is taken
bi-Maxwellian  with $\beta_\| = 1$ and
$T_\perp/T_\|= 1.857$,  thus corresponding to an
initial distance to threshold  $\Gamma_0 \equiv \beta_\perp (T_\perp/T_\| -1)- 1=0.6$.
The retained direction is that of the most unstable mirror mode ($\theta= 72.8^\circ$,
for which the linear growth rate is $\gamma= 5 \times 10^{-3}  \Omega_p$ where $ \Omega_p$
is the ion gyrofrequency). Figures~1 and 2 show that the saturation of the linear instability
leads to the formation of magnetic humps  that, as time elapses, undergo a coarsening process which
leads to their strengthening. As a consequence of the large extension of the
computational domain, a large number of Fourier modes are excited in this simulation. In
agreement with the evolution of the structures, the skewness of the magnetic fluctuations rapidly
becomes positive, increasing up to the moment when the energy of the magnetic fluctuations
saturates. An interesting issue concerns the instantaneous distance $\Gamma$ to threshold  defined by
the left hand side of
Eq.~(\ref{inst-gen}), that rapidly departs from the bi-Maxwellian expression, indicating a significant
evolution of the functional form of the distribution function. It also does not relax to zero but
reaches negative values, while the energy and the skewness
of the fluctuations are still growing under the effect of the nonlinearities.

The departure from a bi-Maxwellian profile for the ion distribution function,
that develops during the simulation,
is exemplified in Fig.~4 which shows that the space-averaged instantaneous distribution function
$\langle f \rangle$ displays a clear flattening near $v_\| = 0$. This  effect can be
associated with a diffusion process in the velocity space due to quasi-linear effects (see Section~6.4).

It is noticeable that when the same simulation is performed in a much smaller computational domain, 
quasi-linear effects are not significant. In contrast, one observes an oscillation of the energy
of the magnetic fluctuations and also of the skewness with a frequency which is consistent
with the bounce frequency $\omega_{\rm tr}^2 =(1/2) v_{{\rm th }\perp} ^2 k_\|^2 (\delta B/B_0)^2$
of the trapped particles, suggesting that particle
trapping is at the origin of these oscillations (which are not present when the computational
domain is large enough).

Similar PIC simulations were previously performed at larger distance from threshold (for which
a lower resolution is sufficient) \cite{Baum03}. They  also clearly indicate that an initial
noise in a mirror unstable regime leads to the formation of magnetic humps whose
number decreases as time elapses (Fig.~4). In addition, these simulations show that in a plasma
with isotropic ion and electron temperatures (thus mirror stable), a localized magnetic
perturbation in the form of a finite-amplitude hump relaxes (Fig.~3),
while a perturbation in the form of a finite-amplitude magnetic hole persists (Fig.~1).

As already mentioned, Eulerian simulations are also reported in \cite{Calif08} in the case of a
relatively small domain ($30 \pi$ ion inertial lengths). In the supercritical regime
(cold electrons and initial bi-Maxwellian ion distribution characterized by $\beta_\| = 6$ and
$T_\perp/T_\| = 1.25$, together with an angle $\theta =83.86^\circ$), the nonlinear development
of the mirror instability leads to the formation of a magnetic hump anticorrelated
with a density hole (Fig.~7). A similar evolution is shown in Fig.~8 at larger distance
from threshold ($\beta_\| = 15$, $T_\perp/T_\| = 1.4$ and $\theta = 78.53^\circ$). Oscillations of the
magnetic-energy fluctuations are visible (as in the PIC simulations in small domains),
that are believed to be  the signature of particle trapping.

Below threshold (isotropic ion temperatures, $\beta_\| = 6$ with $\theta= 83.82^\circ$), an initial
magnetic hole with no density perturbation (Fig.~15) evolves slightly  but
persists, while an anticorrelated density hump develops. Remarkably, a similar evolution takes
place above threshold, as
exemplified in Fig.~13 with $\beta_\| = 6$, $T_\perp/T_\| = 1.36$ and $\theta = 83.82^\circ$.
It is of interest to note that,  above threshold, the magnetic profile develops
overshoots at the boundary of the depression, as in space plasma observations.

In the above simulations, initial magnetic holes are shown to be persistent, but their formation starting from an
initial noise is a different issue.  Two mechanisms were reported for the development of such structures.
In \cite{Calif08}, a PIC simulation is presented (Fig.~16) in a domain
extending on 1024 ion inertial lengths, in conditions relatively far above the mirror instability
threshold ($\beta_\| = 1$, $T_\perp/T_\| = 4$ with $\theta= 50.5^\circ$, leading to a
linear growth rate of $0.156\,  \Omega_p$). At early time, magnetic humps are formed
as in the previously described simulations. However,
later on, these structures progressively evolve towards magnetic holes. The
instantaneous distance $\Gamma$ to threshold decays but now remains slightly positive. The system
is thus continuously stirred  and the coarsening is less efficient. In particular, there is no
formation of isolated structures. Furthermore, the energy of the magnetic fluctuations does not grow monotonically up
to its saturated value  but reaches a sharp maximum before decaying to an equilibrium value (smaller than
the maximum by more than a factor 2). The skewness has a similar evolution with the same time scale.
It reaches a (positive) maximum at the same moment as the energy, and then decays to a negative
value. It is noticeable that in this simulation, the ion distribution function remains relatively close
to a bi-Maxwellian with no significant flattening near the zero parallel velocity.
It turns out that the  transition of the magnetic structures
from humps to holes does not take place with a larger $\beta_\|$ (e.g.,
$\beta_\| = 2$). Another scenario for hole formation is reported in \cite{Genot09} and concerns
a PIC simulation in an expanding domain modeling the magnetosheath \cite{Trav07}. Plots
of the skewness versus $\beta_\|$ and versus the (bi-Maxwellian) distance to threshold
$\Gamma^*= \beta_\perp (T_\perp/T_\| - 1)- 1$ as the system evolves, are presented in Fig.~9 of
\cite{Genot09}. Magnetic humps evolve into holes as they are convected
closer to the magnetosphere within the plasma depression layer where the plasma expends and
its properties change drastically. In particular $\beta_\|$ and $\Gamma^*$ decrease, the transition
from humps to holes occurring for $\Gamma^* \approx 1.7$.

From the combined views of observations and models, the following scenario is proposed in \cite{Genot09}
for the evolution of the mirror structures in space plasmas. ``Large peak structures grow out from moderately
unstable plasma (typically observed behind the bow shock), reaching saturation at sufficiently
large distance from threshold (in the middle magnetosheath depending on the
convection time), and, as $\beta$  is further decreased closer to the
magnetopause, the plasma turns to be mirror stable, which
is accompanied by a decay of the peaks to the profit of hole
structures which can survive these conditions.'' 

\section {Toward a nonlinear theory near threshold}

\subsection{A reductive perturbative expansion}

A natural way for extending the linear theory to the weakly nonlinear regime that develops
near threshold is to perform a long-wavelength reductive perturbative expansion,
based on the observation that near threshold the unstable modes are localized at large scale
\cite{KPS07a, Calif08}.

From the Vlasov equation, one derives the equation for the proton velocity
\begin{equation}
\frac{\mathrm{d}{\pmb u}}{\mathrm{d}t} + \frac{1}{\rho} {\pmb \nabla} \cdot {\bf p}
- \frac{e}{m_p}({\pmb E} + \frac{1}{c} {\pmb u} \times {\pmb B}) = 0, \label{ui}
\end{equation}
where, for cold and massless electrons,
$\displaystyle{{\pmb E} =- \frac {1}{c} \Big ({\pmb u} -\frac{{\pmb j }}{ne} \Big )
\times {\pmb B}}$,
with ${\pmb j} = (c/4\pi) {\pmb \nabla} \times {\pmb B}$.
The ion pressure tensor is rewritten as the sum of gyrotropic and non gyrotropic
(often called gyroviscous) contributions
${\bf p} = (p_{\perp}{\bf n} +
 p_{\|} {\pmb \tau}) + {\pmb \Pi}$,
with  ${\bf n} =  {\bf I} - {\widehat {\pmb b}}\otimes {\widehat {\pmb b}}$  and
${\pmb \tau} = {\widehat {\pmb b}}\otimes {\widehat {\pmb b}}$,
where
${\widehat {\pmb b}}={\pmb B}/|B|$ is the unit vector along the local magnetic field.
Equation (\ref{ui}) is then rewritten in the form
\begin{equation}
\rho \frac{\mathrm{d}{\pmb u}}{\mathrm{d}t}=
-{\pmb\nabla} \Big( p_{\perp}+\frac{|B|^2}{8\pi}\Big )
+ \Big (1+ \frac{4\pi}{|B|^2} (p_\perp -p_\|)\Big )
\frac{{\pmb B}\cdot {\pmb \nabla} {\pmb B}}{4\pi} 
-  \widehat{{\pmb b}} \frac{|B|^2}{4\pi} ({\widehat {\pmb b}}\cdot
{\pmb \nabla} )
\Big ( 1 + \frac{4\pi}{|B|^2}(p_\perp - p_\|) \Big ) -
{\pmb \nabla} \cdot {\pmb \Pi}. \label{NS}
\end{equation}
Only the  projection on the plane perpendicular to the ambient field, obtained  by contracting
Eq.~(\ref{NS}) with the tensor ${\bf n}$, is actually needed. In  order to address
the asymptotic regime, assuming the ambient field in the $z$ direction,
we rescale the independent variables as prescribed by the linear growth rate,
in the form $X= \sqrt \varepsilon x$,
$Y= \sqrt \varepsilon y$, $Z= \varepsilon z$, $T = \varepsilon^2 t$, where $\varepsilon$
measures the distance to threshold. We furthermore expand any field $\varphi$ in the form
\begin{equation}
\varphi = \sum_{n=0}^\infty \varepsilon^{n/2} \varphi_{n/2}.
\end{equation}
In particular
\begin{eqnarray}
&&{\pmb B}_\perp = \varepsilon^{3/2} {\pmb B}_\perp^{(3/2)} +  \varepsilon^{5/2} {\pmb B}_\perp^{(5/2)}
+ \cdots\\
&&B_z = B_0 + \varepsilon B_z^{(1)} + \varepsilon^2  B_z^{(2)} + \cdots.
\end{eqnarray}
 From the Faraday equation and the assumption of
 cold and massless electrons that implies  ${\pmb E}\cdot {\pmb B}=0$, one has for the electric field
\begin{eqnarray}
&&{\pmb E}_\perp = \varepsilon^{5/2} {\pmb E}_\perp^{(5/2)} +  \varepsilon^{7/2} {\pmb E}_\perp^{(7/2)}
+ \cdots\\
&&E_z = \varepsilon^4  E_z^{(4)} + \varepsilon^5  E_z^{(5)}\cdots.
\end{eqnarray}
One shows that the parallel current vanishes to leading order ($ {\pmb \nabla}_\perp \times {\pmb B}_\perp^{(3/2)} =0 $)
which, when combined  with the
divergenceless condition $\displaystyle{ {\pmb \nabla}_\perp \cdot  {\pmb B}_\perp^{(3/2)} +
\partial_Z B_z^{(1)} = 0}$, implies
${\pmb B}_\perp^{(3/2)} = (-\Delta_\perp)^{-1}  {\pmb \nabla}_\perp \partial_Z B_z^{(1)}$.

The ion velocity equation is then rewritten in the form
of the pressure balance equation
\begin{equation}
{\pmb \nabla}_\perp \Big [ {\overline p_\perp} + \frac{B_0}{4\pi}b_z
+ \varepsilon \frac{b_z^2}{8\pi} +
 \frac{2\varepsilon}{\beta_\perp} \Big (1 + \frac{\beta_\perp -\beta_\|}{2}
\Big )p_\perp^{(0)} (\Delta_\perp)^{-1} \partial_{ZZ} \frac{b_z}{B_0} \Big ]
+ \varepsilon \Big ( {\pmb \nabla} \cdot {\pmb \Pi} \Big )^{(5/2)}_\perp = O(\varepsilon^2), 
\label{equil}
\end{equation}
where, because of the cancelation of the leading order at threshold, we have
defined $b_z= B_z^{(1)}+ \varepsilon B_z^{(2)}$ and ${\overline p_\perp} = {p_\perp}^{(1)} +
\varepsilon p_{\perp}^{(2)}$.
In this regime, the ion inertia term is subdominant, and the time dependency
originates from the Landau resonance through the transverse pressure.
This contrasts with the MHD approach where it arises from the advection term.

In Eq.~(\ref{equil}), the perpendicular pressure and the contribution of the
gyroviscous tensor gradient are to be computed from the proton distribution function
$f = f^{(0)} + \varepsilon f^{(1)} + \varepsilon^{3/2} f^{(3/2)}+ \varepsilon^{2} f^{(2)}+ \cdots$,
where the leading order is the equilibrium state  and the other contributions are obtained by
perturbative resolution of the Vlasov--Maxwell equations. Detailed computations are presented in
\cite{Calif08} in the case of a bi-Maxwellian equilibrium state and in \cite{HKP09} for
an arbitrary equilibrium distribution function. After returning to the unscaled variables, this
leads to the dynamical equation
\begin{equation}
\partial_t b
={\sqrt{\frac{2}{\pi}} \tilde{v}} \left(-{\cal H} \partial_z \right)
\left( \Gamma b + \frac{3}{2}{\tilde r}^2 \Delta_\perp b
 - \chi \partial^2_z\Delta_\perp^{-1} b -  \Lambda b^2 \right),\label{kuznplus}
\end{equation}
that involves several coefficients (and among them the distance to threshold $\Gamma$)
whose expressions in terms of
momenta of the equilibrium distribution and of its derivatives,
are given in Ref.~\cite{HKP09}. Here $\cal H$ denotes the Hilbert transform and the
operator ${\cal K}_z =-{\cal H} \partial_z$ has the Fourier symbol $|k_z|$.
In the case of a bi-Maxwellian distribution function, Eq.~(\ref{kuznplus}) reduces to
\begin{equation}
\partial_t b =
\sqrt{\frac{2}{\pi}}v_{\rm th\, \|}\frac{\beta_\|}{\beta_\perp^2} \Big (- {\cal H}
\partial_z \Big )  \Big \{ \Big [\beta_\perp \Big (\frac{\beta_\perp}{\beta_\|} -1 \Big ) - 1 \Big ] b
+ \frac{3}{2}r_{\rm L}^2 \Delta_\perp b 
-\Big (1 + \frac{\beta_\perp - \beta_\|}{2}  \Big )
\Delta_\perp^{-1}\partial_{zz} b - \frac{3}{2}
\Big ( \frac{1+ \beta_\perp}{\beta_\perp^2} \Big ) b^2 \Big \},\label{kuzn}
\end{equation}
Equations~(\ref{kuznplus}) or (\ref{kuzn}) can be viewed
as an extension of the linear dispersion relation  by a nonlinear contribution that at the order of
the expansion is not affected by FLR corrections, and can thus be estimated in the framework of
the drift kinetic approximation \cite{KPS07a, Pokho08}. After a simple change of variables,
Eq.~(\ref{kuznplus}) becomes
\begin{equation}
\partial_\tau u = (- {\cal H} \partial_\xi )   (\sigma u  + \Delta_\perp u -
\Delta_\perp^{-1}\partial_{\xi \xi} u - 3 \, {\rm sgn} (\Lambda)\,  u^2),\label{model0}
\end{equation}
where $\sigma ={\rm sgn} \, \Gamma$ is equal to $1$ above threshold and $-1$ below threshold. Note
that changing the sign of the nonlinear coupling coefficient $\Lambda$ leads to replacing $u$
by  $-u$ and thus magnetic  holes by magnetic humps
or  the reverse. For a bi-Maxwellian equilibrium, $\Lambda>0$. In this case, when
looking for solutions depending on a unique space coordinate along a prescribed direction quasi-transverse
to  the ambient field, Eq.~(\ref{kuzn}) reduces to
\begin{equation}
\partial_t u = {\cal K}_\Xi [ (\sigma u  + \partial_{\Xi\Xi})\ u - 3  u^2)]. \label{model1}
\end{equation}
A numerical integration of Eq.~(\ref{model1}) above
threshold with initial conditions in the form of a sinusoidal function of weak amplitude displaying
several wavelengths in the integration domain, is presented in Ref.~\cite{KPS07a}. After the early
linear phase, several magnetic holes are formed, whose number is progressively
reduced to one. Nevertheless, the asymptotic equation cannot capture the nonlinear saturation
of the instability and, after a while, the solution blows up, indicating the
breakdown of the asymptotic scaling and the formation of large-amplitude structures.
Near collapse, the solution displays a self-similar profile discussed in  \cite{KPS07a}. Furthermore,
as detailed in Ref.~\cite{KPS07b},  Eqs.~(\ref{model0}) and (\ref{model1})
do not admit stationary non-zero solutions above threshold, while below threshold steady
solutions (in the form of magnetic holes) exist but are unstable. It indeed  turns
out that the instability is associated with a subcritical bifurcation, a property consistent with
the bistability property reported from both satellite observations and numerical simulations.

Nevertheless, the reductive perturbative expansion near a bi-Maxwellian equilibrium state,
that predicts the formation of large-amplitude magnetic holes above threshold, contradicts
the results of direct numerical integrations of the Vlasov--Maxwell equations
with similar initial conditions, which lead to magnetic humps.
The question thus arises of the origin of this discrepancy which could be the consequence of one
of the two main assumptions involved in the asymptotic theory, namely the fact that the ion
distribution function remains close to a bi-Maxwellian, and also that kinetic effects are retained
at the linear level only.
In addition, as already mentioned, this theory  does not capture the saturation
of the instability which is in fact  strengthened by the retained nonlinearity.
Before revisiting the above theory, let us briefly review phenomenological models 
for the  saturation of the mirror instability, previously discussed in the literature.

\subsection{Previous phenomenological modeling of the saturation}.

The models proposed in \cite{pant95, kiso96, pant98} are mostly based on a separation between
trapped (with large pitch angle) and untrapped (with small pinch angle) particles
which respond differently to magnetic field variations,
the saturation of  the instability resulting from
the cooling of trapped   particles in magnetic holes.
In the rising field regions, trapped particles are indeed excluded by the mirror
force, leading to a decrease of the particle pressure (reduction of $\beta_\perp$)
and to the evolution to marginal stability (without important change  in the
particle energy). In the well regions, no particle can be excluded.
Some trapped particles are cooled by loosing perpendicular energy,
leading to a reduction of the temperature anisotropy.
Large reductions in the field are required in the wells in order to
cool the trapped population enough to stabilize the system.
Such models mostly predict deep magnetic fields in conditions of marginal stability.
They hardly explain the formation of magnetic humps
(except possibly for unusually  high $\beta$'s), and do not
address the phenomenon of bistability.

Particle trapping is difficult to retain within a systematic reductive perturbative analysis.
Within this formalism, it can nevertheless be phenomenologically described by a renormalization
of the time derivative, on the basis that it produces a quenching of the Landau resonance.
This argument is quantitatively implemented  in \cite{Pokho08} by prescribing a flattening
of the distribution function of the parallel velocity on a range that extends with the amplitude of
the magnetic perturbation. This leads the authors to modify Eq.~(\ref{model1}) by multiplying the term
$\partial_t u$ by a nonlinear function of $u$ that prevents the explosive growth of the amplitude and,
by vanishing, yields stationary solutions in the form of negative  Korteweg--de Vries solitons. Yet,
only magnetic holes can result from this model.

\subsection{Modeling the variation of the local ion Larmor radius}

As already mentioned, either the persistence to leading order of a bi-Maxwellian distribution function or the
linear description of the kinetic effects (in particular the FLR corrections),
that both result from the reductive perturbative expansion scalings, are possibly at the origin
of the discrepancy with the results of
the slightly supercritical direct numerical integrations  of the Vlasov--Maxwell equations
with an  initial bi-Maxwellian ion distribution function.

The model presented in \cite{KPS07a} consists in modifying
Eq.~(\ref{model1}) in order to  phenomenologically take into account
the variation of the local ion Larmor radius due to the development of magnetic structures. Indeed, in regions of
weaker magnetic field, the ion Larmor radius is larger, making stabilizing effects of FLR
corrections more efficient
than in the linear regime. Consequently, the mirror instability is more easily quenched in magnetic field
minima, making magnetic humps more likely to form in the saturating phase of the mirror instability.
The conservation to leading order of the magnetic moment suggests that the square Larmor radius scales like the inverse
longitudinal magnetic field, which leads  to replace the FLR contribution $\Delta_\perp u$ in Eq.~(\ref{model0})
by the nonlinear term ${\frac{1}{1+\alpha u} \Delta_\perp u}$ (possibly supplemented by a higher order
correction in the form of a bi-Laplacian to improve accuracy). Here the parameter $\alpha$ is given by
$\alpha = \frac{2\beta_\perp}{1 + \beta_\perp} \Big |\beta_\perp \Big (\frac{T_\perp}{T_\|} -1 \Big ) -1 \Big |$
and is thus proportional to the  (unsigned) bi-Maxwellian distance to threshold.
 Numerical integrations of the
resulting equations are presented in \cite{Calif08}.
Figure 10 of this reference shows the evolution of
the solution from an initial noise in the supercritical regime ($\alpha=1.54$). The singularity is arrested,
the linear instability saturating
with the formation of magnetic humps whose number decreases in time by an effect of coarsening that tends
to strengthen the mirror structures. Very close to threshold ($\alpha=0.05$), the system evolves to a regime
of ordered fluctuations with a slight dominance of the minimum amplitude. Below threshold (but not
too far in the subcritical regime), large initial perturbations lead to the formation of
magnetic holes.
Figure 11, that displays the skewness of the solution versus $\sigma \alpha$ (a measure
of the signed distance to threshold weighted by a function of  $\beta_\perp$), is
qualitatively consistent with the observational data: the skewness is negative below
threshold and becomes positive sightly above threshold.

\subsection{Modeling quasi-linear effects}

An important observation from  the numerical simulations of the Vlasov--Maxwell equations near
threshold concerns the flattening
of the ion distribution function near the zero parallel velocity (Fig.~4 of \cite{Calif08}), that results from a
diffusion in the velocity space possibly described by the quasi-linear theory. This approach 
(see e.g. \cite{DII10}, Section 3.1) was in fact early
applied to the mirror dynamics \cite{shsh64}. It assumes space homogeneity and thus the absence of coherent
structures. It can thus be valid at early times only. It requires also many modes in interaction, and
consequently  an extended domain. Furthermore,
the extension of the quasi-linear theory to an aperiodic instability is not straightforward,
but turns out to be justified in the case of the mirror instability near threshold because of its small
growth rate ($\gamma_k \ll k_z v_{\rm th \|}$). The method  can be sketched as follows.
Let the instantaneous ion distribution function be written in the form
$f({\pmb x},{\pmb v},t)  = f_0({\pmb v},t) + f_1({\pmb x},{\pmb v},t)$ where
$f_0({\pmb v},t) = \langle f({\pmb x},{\pmb v},t) \rangle_ {\pmb x}$
corresponds to the space-averaged contribution.
The perturbation $f_1$ is decomposed in Fourier modes in the form
$f_1 = \frac{1}{2} \sum_{\pmb k} f_{\pmb k} \, \mathrm{e}^{\mathrm{i}{\pmb k} \cdot {\pmb x}}$. Similarly, the electric field reads
${\pmb E } =  \frac{1}{2} \sum_{\pmb k} {\pmb E}_{\pmb k} \, \mathrm{e}^{\mathrm{i}{\pmb k} \cdot {\pmb x}}$. The Faraday equation
prescribes the fluctuating Fourier modes in the form
${\pmb B}_{\pmb k}= -\frac{c} {\omega_{\pmb k}} {\pmb k} \times  {\pmb E}_{\pmb k}$.
Substituting in the Vlasov equation, one gets
\begin{equation}
\partial_t f_0 = - \frac{q}{2m} \sum_{\pmb k} \Big \{ {\pmb E}_{\bf k} ^* \Big (1 - \frac{{\pmb k}
\cdot {\pmb v}}{\omega_{\pmb k}^*}\Big )
+ \frac{{\pmb k}}{\omega_{\pmb k}^*} ({\pmb v} \cdot {\pmb E}_{\pmb k}^*)\Big \}
{\pmb \nabla}_{\pmb v} f_{\pmb k} + {\rm c.c.}
\end{equation}
A main hypothesis is that the variation of $f_0$ is slow compared with that of the fluctuations.
Furthermore, the amplitude of the harmonic modes $f_{\pmb k}$ are assumed to be of small amplitude,
and consequently their interactions are neglected.
In this framework, the modes $f_{\pmb k}$ are expressed as linear functionals of $f_0$. After some algebra, this
leads to a diffusion equation in velocity space for the space-average distribution function that, when
suppressing the subscript zero in order to simplify the writing, takes the form
\begin{equation}
 \frac{\partial  f}{\partial t} =
\frac{\partial\ }{\partial v_\parallel} D_{\|\|} \frac{\partial  f }{\partial v_\parallel}
+ \frac{1}{v_\perp} \frac{\partial }{\partial v_\perp} v_\perp
\left( D_{\perp\|} \frac{\partial  f }{\partial v_\parallel} + D_{\perp\perp}
\frac{\partial  f }{\partial v_\perp} \right)
\label{qlind}
\end{equation}
with
\begin{equation}
D_{\|\|}  = v_\perp^4  \sum\limits_{\pmb k}
\frac{|b_{\pmb k}|^2}{4}
\frac{ \gamma_{\pmb k} k_\parallel^2 }
 {k_\parallel^2 v_\parallel^2+\gamma_{\pmb k}^2}, \ \ \
D_{\perp\|} = -2 \frac{v_\|}{v_\perp} D_{\|\|},\ \ \
D_{\perp\perp} = v_\perp^2 \sum\limits_{\pmb k}
\gamma_{\pmb k} \frac{|b_{\pmb k}|^2}{4},
\label{D}
\end{equation}
supplemented by ($b_{\pmb k} = \delta B_z ( {\pmb k} )/B_0$)
\begin{equation}
\frac{\partial b_{\pmb k}}{\partial t}= \gamma_{\pmb k} b_{\pmb k},
\label{Bzqlin}
\end{equation}
where $\gamma_{\pmb k}$ is the linear growth rate, computed using Eq. (\ref{rate}) where the coefficients are
evaluated using the instantaneous value of the space-averaged distribution function. 
The resulting  system thus concentrates on the spatially independent part of the distribution function,
ignoring the nonlinearities associated with space variations and describing the wave--wave interactions. It
is solved numerically in \cite{HKP09}. The results are summarized in Fig.~1 of this reference.
It clearly displays
the saturation of the energy of the magnetic fluctuations.  The system is shown to evolve towards a
regime of marginal stability as the parameter $\Gamma$ that measures the instantaneous distance to threshold
decays to zero (an asymptotic value that is actually difficult to reach numerically).
Simultaneously, the growth rate
also decreases to zero. In addition, the gray scale plot of $v_\perp \delta f$ where $\delta f$
denotes the difference of the solution (corresponding to the space averaged distribution function)
in the saturated regime with its initial value (Fig.~2)
accurately reproduces the qualitative features of Fig.~4 of \cite{Calif08}
where the same quantity is evaluated from the direct numerical simulation
of the Vlasov--Maxwell equations in an extended domain near threshold:
both figures display a similar diffusion in the velocity space near the resonance.

\subsection{Coupling quasi-linear theory and reductive perturbative expansion}

The observation that the early nonlinear dynamics of the mirror modes is accurately described by
the quasi-linear theory before  the formation of strong coherent structures, suggests
to couple the quasi-linear theory and the reductive perturbative expansion by estimating the
coefficients entering  Eq.~(\ref{kuznplus}) not from the (initial) equilibrium state but 
from the instantaneous quasi-linear distribution function
\cite{HKP09}, a procedure consistent with the key assumption 
$\frac{1}{\langle f\rangle}\frac{\partial{\langle f\rangle}}{{\mathrm d}t}\ll \gamma_{\pmb k}$
of the quasi-linear theory.
Nevertheless, because of the nearly singular character of the distribution function resulting
from the quasi-linear evolution near the zero parallel velocity, contributions of the resonant
particles are to be taken into account when estimating the nonlinear coupling. This effects,
which corresponds to the nonlinear Landau damping, leads to the denominator in Eq.~(\ref{coupl2})
\begin{eqnarray}
&& \frac{\partial  f}{\partial t} =
\frac{\partial\ }{\partial v_\parallel} D_{\|\|} \frac{\partial  f }{\partial v_\parallel}
+ \frac{1}{v_\perp} \frac{\partial }{\partial v_\perp} v_\perp
\left(
  D_{\perp\|} \frac{\partial  f }{\partial v_\parallel} + D_{\perp\perp}
\frac{\partial  f }{\partial v_\perp} \right)  \label{coupl1}\\
&&\frac{\partial  b}{\partial t}
= \frac{ {\sqrt{\frac{2}{\pi}} \tilde{v}} }{1 + 2 \frac{\tilde v}{v_\Lambda} b}
  \left(-{\cal H} \partial_z \right)
\left( \Gamma b + \frac{3}{2}{\tilde r}^2 \Delta_\perp b
 -  \chi \partial^2_z\Delta_\perp^{-1} b -  \Lambda b^2 \right),\label{coupl2}
\end{eqnarray}
where the additional parameter $v_\Lambda^{-1}$ that measures the
strength of  the resonance, is given as an explicit functional of
the instantaneous distribution function.
The above system can also be viewed as retaining the full nonlinear evolution
for the magnetic fluctuations given by the reductive perturbative theory,
in the estimate of the coefficients of the quasi-linear diffusion equation, 
instead of the sole linear contribution.
The resulting  system was numerically integrated in the supercritical regime
in \cite{HKP09}. One observes that while 
the time evolution of the distance to threshold and of the instantaneous
growth rate are weakly perturbed compared with the genuine quasi-linear theory,
the nonlinear coupling parameter $\Lambda$ and the resonance coefficient
$v_\Lambda^{-1}$ change sign at early time (Fig.~3). Furthermore, after the linear phase,
the local maximum amplitude of the magnetic fluctuations no longer saturate,
but displays a fast growth associated with  the formation of large-amplitude structures.
As previously mentioned, the change of the sign of the
$\Lambda$ coefficient modifies  the shape of the
structures. Indeed, as displayed in Fig.~4, the emerging structures are now magnetic humps,
as observed in the direct numerical simulations of the Vlasov--Maxwell equations.

Capturing the saturation of the mirror structures by integration of Eqs.~(\ref{coupl1})--(\ref{coupl2})
is rather
delicate due to numerical limitations. In order to study the saturation effect produced by the
denominator associated with the nonlinear Landau damping in Eq. (\ref{coupl2}), 
a simplified approach used in \cite{PSKH09}
consists in freezing the coefficients after the early quasi-linear phase present in the supercritical regime,
leaning on the quasi-singular character of the subsequent nonlinear dynamics.
In the subcritical regime, the quasi-linear phase does not occur, but the denominator
in Eq.~(\ref{coupl2}) is nevertheless
retained because of the large amplitude of the fluctuations needed to capture non trivial solutions.
This leads to the model equation
\begin{equation}
\partial_t b = \frac{ 1 }{1 + s\alpha b} (-{\cal H} \partial_\xi)
\left( \sigma b + \mu \partial_{\xi\xi} b -  3s b^2 \right), \label{sat}
\end{equation}
where $\xi$ is the coordinate along the most unstable direction. Here,
$\sigma=\pm 1$ depending on the super or sub-critical regime, and
$s=\pm 1$ depending on the sign of the nonlinear coupling. The latter is
negative above threshold after the early quasi-linear phase,
and positive in the case of persistence
of a bi-Maxwellian distribution as in the subcritical regime.

Numerical integration of Eq.~(\ref{sat}) in the supercritical regime
shows that during the nonlinear phase of amplitude growth, a plateau
of negative values gradually develops, that tends to locally
reduce the ambient magnetic field, putting the system in a
situation similar to the subcritical regime.
The solution is then attracted to a steady state given by
the negative of the Korteveg--de Vries soliton profile and
a maximal amplitude $b_{\rm max} = 1/\alpha$.
The amplitude of the structures is prescribed by the strength of
the early-time quasi-linear  resonance: larger amplitudes are obtained
when these effects are smaller.
When starting with random initial conditions, which leads to a
large number of humps, a coarsening phenomenon is observed.
If in the  supercritical regime quasi-linear effects are subdominant,
$s=+1$ and magnetic holes are obtained.
Below threshold ($\sigma = -1$), the distribution function  remains
bi-Maxwellian ($s = +1$) and large-amplitude magnetic holes  are stable.

The above discussion where the quasi-linear diffusion plays a main
role, assumes an extended system. When because of geometrical
constraints such as a small computational domain,
only a few modes are unstable, particle trapping can play a role but its effect on 
on the distribution function is in fact similar to that of the quasi-linear diffusion,
both processes leading to a flattening near the zero parallel velocity
\cite{Pokho08,IPB09} and consequently to a change of the sign of the nonlinear
coupling.
A possibly more refined description of trapping could consists in
computing the coefficients entering Eq.~(\ref{kuznplus})  on the basis of a distribution
function that distinguishes between free ions, trapped ions and tail particles,
as suggested in the construction of the steady mirror structures described in \cite{JS09}.

\section{Turbulent plasma with mirror modes}

\subsection{Observational results}

Cluster measurements in the Earth's magnetosphere close to the magnetopause, presented
in \cite{Sahra06} concern a regime where the turbulence is compressible, anisotropic and dominated
by mirror structures. Figure 1 of this letter displays high-level quasi-regular oscillations on the
parallel component of the magnetic fluctuations, signature of strong compressible waves.
The so-called k-filtering technique \cite{PL91} enabled the authors to discriminate between
time and space variations. Figure 1 indeed shows a $f^{-7/3}$ range for the low-frequency
temporal spectrum, while Fig.~6 displays a $k_v^{-8/3}$ behavior for the spatial spectrum
in the direction of the flow (referred to by the index $v$).  It turns out that the ``waves'' are actually
stationary structures in the flow frame, identified as mirror modes (see Fig.~3 and also
\cite{Walker04, Lucek05}).
It appears that energy is injected at large scales via a mirror
instability well predicted by the linear theory, and cascades nonlinearly more preferentially
along the flow direction (see Fig.~5), down to scales of the order of a few ion Larmor radii.

The values of the proton temperature anisotropy and of the parameter of $\beta_{\|}$ measured
in the slow solar wind by the WIND/SWE mission  are considered in \cite{Hell06},
and their relation with the temperature-anisotropy-driven instabilities
are discussed. A main result is that the
proton temperature anisotropy is constrained by oblique instabilities, either mirror or firehose,
depending on the dominance of the perpendicular or parallel temperature respectively. This observation
contrasts with  the linear theory which predicts a dominance of the proton cyclotron and of the
parallel fire-hose instabilities. It is exemplified in Fig.~1 of \cite{Hell06}
that shows a color-scale plot
of the relative frequency of ($\beta_{\| p}$, $T_{\perp p}/ T_{\| p}$) for the considered data,
together with contours of estimated growth rates of the above instabilities 
in a plasma with  bi-Maxwellian ions and isothermal electrons. A similar conclusion is reached in \cite{Bale09}
by analyzing $10^6$ independent measurements of gyroscale magnetic fluctuations in the solar wind.
An additional result is that these fluctuations are enhanced along the temperature threshold of the mirror,
proton oblique firehose, and ion cyclotron instabilities. In addition, the relative importance of
the parallel magnetic fluctuations, as measured by
$|\delta {\pmb b}_\||^2/(|\delta {\pmb b}_\||^2 + |\delta {\pmb b}_\perp|^2)$,
is increased  at $\beta_\| \gtrsim 1$ along the mirror threshold but small
elsewhere, as expected for a mirror instability.

The question then arises of the origin of the
temperature anisotropy. As already mentioned, in the planet magnetosheaths
large-perpendicular-to-parallel ion temperature anisotropy ratio are due to  compression downstream
the quasi-perpendicular shock or to the flow of plasma with high parallel velocity close to the magnetopause.
In the solar wind, the mechanism that is usually believed to drive such an anisotropy is the proton resonance
with left-handed cyclotron waves. However, as recently noticed in \cite{Bourou10}, this process requires
a parallel cascade via parametric decay to high frequency modes, while the development of quasi-perpendicular
low-frequency kinetic Alfv\'en waves is usually considered
to be the dominant process.  The statistical analysis presented in \cite{Bourou10} of data
in the fast solar wind provided by
Helios 2 primary mission at radial distance from the sun from $0.3$ to $0.9$ AU (where $\beta_\|$ 
ranges between about $0.1$ and $0.6$),
shows a correlation between enhancements of the power density of transverse waves
(with frequency between $0.01$ and $0.1$ ion gyrofrequency in the plasma frame)
and increases of the ratio of the perpendicular-to-parallel proton temperatures. This led the authors to conclude
that perpendicular ion heating is connected with the presence of a
large-scale turbulence where transverse oscillations are dominant.

\subsection{Numerical simulation of the mode competition}

It has often be argued \cite{Gary92, Gary93} that, in a plasma with a dominant perpendicular temperature,
the presence of  ${\rm He}^{++}$  which lower the
linear ion-cyclotron growth rate without significantly affecting the mirror modes, could be at the origin of the
dominance of the latter structures in planetary magnetosheaths. Such an ambiguity associated with
multiple ion effects is not present in numerical simulations where an electron-proton plasma can be considered.
Such simulations were performed in \cite{Shoji09} using hybrid PIC simulations in two and three space  dimensions.
Figure 2 of this paper shows that in the linear regime (see also \cite{GaryBook93}), the ion-cyclotron (referred to
as L-mode EMIC) growth rate dominates that of the mirror modes. This ordering does not necessarily persist
in the non linear regime. To address this question, the authors
identify the magnetic energy density of each wave by the following method.
Examination of the wave spectra shows that ion cyclotron waves and mirror mode waves are separated by
the angle $\theta = 30^\circ$  between the wave vector ${\pmb k}$ and ${\pmb B}_0$. Accordingly,
the ``L-mode EMIC wave range'' is assumes to be defined by $0^\circ \leqslant \theta  \leqslant 30^\circ$,
and the ``mirror mode wave range'' by $30^\circ < \theta \leqslant 80^\circ$. The spectra of the magnetic field
is divided into two ranges of the wave vector space, and the energy associated with the two
types of waves is obtained by
integrating the square magnetic modes $|{\pmb B}({\pmb k})|^2/|B_0|^2$ in each of these ranges. Figure 6
compares the time evolution of the energy associated with the two types of waves, both in two and three dimensions.
It appears that in two dimensions, the mirror energy is strongly subdominant. Differently,  in three
dimensions the two energies reach the same maximum after about the same time but, while the mirror energy then remains
quasi-stationary, the ion-cyclotron energy rapidly decays in time making the mirror modes strongly dominant in the
long-time regime.

\subsection{Toward fluid simulations of the mirror dynamics}

Strictly speaking an accurate description of collisionless plasmas even in the presence of a
strong ambient field requires a fully kinetic description, but in the turbulent regime
huge computational resources are requested for such simulations.
On the other hand, it is clear that usual MHD or bi-fluid descriptions are unable to
accurately describe such a regime, as exemplified by the mirror instability.
It is thus of interest to  extend the fluid approach in order to include  leading
low-frequency kinetic effects such as Landau damping and FLR corrections.
This approach was initiated in \cite{HP90} where
a closure relation retaining linear Landau damping was  suggested. It
was implemented in the case of large-scale
MHD in \cite{SHD97}, as a closure of the moment hierarchy in
the framework of the kinetic drift approximation, leading to a so-called Landau fluid model. It was extended to
dispersive MHD with Hall effect and large-scale  FLR corrections in \cite{PS03, GPS05}.
A generalization of the model including a description of quasi-transverse scales significantly smaller than
the ion gyroscale, under the gyrokinetic scaling,
is presented in \cite{PS07} and referred to as FLR-Landau fluid model.
Note that in contrast with the gyrokinetic description,
Landau fluids retain fast waves that are accurately described up to the
ion gyroscale. Nevertheless, like the gyrofluids,
they neglect wave--particle trapping, that is to say the effect of
particle bounce motion on the distribution function near
resonance.

We cannot detail here the derivation of the FLR-Landau fluid model nor even write the full equations explicitly.
In brief, it consists in deriving from the Vlasov--Maxwell
equations for the protons and the electrons a hierarchy of exact equations for the plasma density and velocity
(for the sake of simplicity the electron inertia can be neglected, leading to a generalized Ohm's law),
and for the  gyrotropic pressures and heat fluxes of each particle species.
The closure concerns the gyrotropic components of the fourth order cumulants of each species
and the non gyrotropic components (FLR corrections)
of the  pressures, heat fluxes and fourth order moments of the ions  (electron FLR corrections are neglected,
as the description does not go beyond the ionic scales).
It is built in such a way that the linear kinetic theory
is reproduced in the low-frequency limit, within the gyrokinetic scaling
in the quasi-transverse directions but also
at all the hydrodynamic scales. For this purpose, the various moments are
first computed within the usual linear kinetic theory in the above
asymptotics, leading to expressions in terms of the electromagnetic
fluctuations. The difficulty is that these formulae  involve
the electrostatic response function (or the plasma dispersion function) that
depends on the phase velocity of the corresponding
Fourier mode and consequently  cannot be implemented in the framework of an initial value
problem. The idea is first to combine these
expressions for  the various moments in order to express the unknown cumulants in terms of lower order moments and
of the magnetic current. In such expressions, the response function arises within a rational
function. At this step, the response
function is replaced by a Pad\'e approximant whose number of poles is
chosen in such a way that, when returning to the physical space,
the closure relations reduce to  partial differential equations, usually of the first order
in time. They involve an Hilbert
transform in the longitudinal space variable, signature of the Landau resonance. Being derived
in the context of the
linear kinetic theory, these closures involve equilibrium quantities such as the parallel and perpendicular pressures
or temperatures.
In order to take into account the global evolution of the plasma, such as its global heating,
it is suitable to replace these equilibrium (i.e., initial) quantities by the instantaneous space-averaged values.
A similar approach is implemented to evaluate the relevant FLR corrections. A detailed derivation
of the model is presented in
\cite{PS07}. Explicit expressions for the various contributions in the one-dimensional regime
where the space variations only take place in a direction making a
prescribed angle with the ambient field, are given in \cite{BPS07}.

Detailed comparison with the kinetic theory
demonstrates that the FLR-Landau fluid model correctly reproduces the mirror instability
and its quenching at small scale (Fig.~10 of \cite{PS07}). It also accurately reproduces
the frequency and the damping rate
of  the (quasi-transverse) kinetic Alfv\'en waves (Fig.~2). In oblique directions (Fig.~3), the accuracy is
limited by the presence of resonances.

The nonlinear dynamics of mirror modes in one space dimension is studied using the FLR-Landau fluid in \cite{BPS07}.
Compared with Eq.~(\ref{model1}) which results from the reductive perturbative expansion of the
Vlasov--Maxwell equations, the FLR-landau fluid model  involves richer nonlinearities.
As a consequence, it is capable to arrest the singularity.
Nevertheless, in both models, the distribution function
remains close to a bi-Maxwellian and FLR corrections are described at an essentially linear level.
As a consequence, in both cases, the saturation of the mirror instability leads to
magnetic holes and not humps. Figure~2 of \cite{BPS07} exemplifies in the FLR-Landau fluid framework
the development of the mirror instability and the formation of
magnetic holes  anticorrelated with density humps, and their slow relaxation towards the uniform state
on a long  time scale of the order of $10^5 \,\Omega_p^{-1}$, with
the temperatures evolving toward a mirror stable regime (Fig.~3). In order to prevent this
relaxation, simulations were also performed when the mean ion pressures are prescribed to
constant values. In a computational domain of relatively small extension,
the system evolves to a stationary state (Fig.~4).
Through the magnetic holes, the increase of the perpendicular temperature is larger than that of the
parallel one. It is of interest to note the agreement with observations reported in \cite{Tsuru02}
that solar wind protons detected within magnetic holes  are found to be
preferentially heated perpendicular to the ambient field. Simulations of the FLR-Landau fluid with prescribed mean
pressures in an extended domain and at moderate distance from threshold (Fig.~5) display a regime of
spatio-temporal chaos. Holes form and disappear in an erratic fashion, and large-scale compression waves are
observed to propagate through the domain.

Furthermore, below the mirror instability threshold, large-amplitude stationary solutions have been obtained using
a continuation procedure (Fig.~6). Moreover, the solution displays  a small magnetic field component
perpendicular to the $({\pmb k}, {\pmb B_0})$-plane (Figs.~7 and 8). Interestingly, this component is
symmetric with respect to the center of the magnetic hole, a property also observed in direct Vlasov--Maxwell simulations
(\cite{Calif08}, Fig.~13),
which contrasts with soliton models based on Hall-MHD \cite{Sta04a, Sta05, Mjolhus06, Mjolhus09}.
All the velocity components are anti-symmetric, while the component perpendicular to the $({\pmb k}, {\pmb B_0})$-plane
has a relatively large amplitude with a sharp gradient at the center of the hole (Fig.~8),
indicating that the gyroviscous tensor plays an important role in the
equilibrium. Similar signatures are observed in PIC simulations of
non-propagating rarefractive solitary structures generated by
particle injection \citep{BSD05} in a plasma that is initially isotropic.
The heat fluxes also seem to play an important
role since no equilibrium can be reached when these terms are removed from the equations.

Although the FLR-Landau fluid model does not accurately reproduce the geometry of the nonlinear mirror modes
in the weakly unstable regime, the influence of the mirror instability on the
development of the temperature anisotropy of a turbulent plasma can nevertheless be correctly captured. In this
context, it is of interest to mention recent simulations performed in one space dimension along a direction
quasi-transverse to  the ambient field \cite{LMPS10}. A turbulent regime is considered where the system is
driven by a random force acting when the sum of the kinetic and magnetic energies is smaller than a
prescribed value (changing this value  mostly affects the evolution time scale),
in a way aimed to create large-scale transverse oscillations. It turns out that for $\beta_\|$ larger than a critical
value close to  $0.35$, a temperature anisotropy  corresponding to a
dominant perpendicular ion heating develops. As time elapses, the system
reaches a regime where the mirror instability  develops and constrains the anisotropy at a level
slightly in excess of the threshold condition estimated by Eq.~(\ref{electrons}). In the physical space, mirror
structures are clearly identified. At smaller $\beta_\|$, the parallel ion temperature dominate, leading to
significantly different dynamics. More detailed analyses of these runs are in progress, but the present
simulations already provide
an interesting insight on the role of turbulence and non-resonant interactions
on the temperature anisotropy observed in the solar wind, whose origin is still debated
\cite{Bourou10, Bourouaine08,Chandran10}.

\section{Summary }

We have presented an overview of the present understanding of the dynamics of the
mirror modes commonly observed in space plasmas displaying a dominant ion perpendicular temperature
and a relatively large $\beta$. Various questions arise, originating in particular
from the existence of mirror structures in plasmas that are stable for the mirror
instability, and also  from their appearance as  both magnetic enhancements or depressions,
with a statistical predominance of the latter. A main conclusion resulting from
recent analyzes of the mirror data provided by satellite missions is that magnetic humps are
only observed in an unstable plasma, while in the mirror-stable regime only magnetic
holes are present. Direct numerical simulations of the Vlasov--Maxwell equations for
initially bi-Maxwellian ions (and cold electrons) display
a flattening of the ion distribution function near the zero parallel velocity,
viewed as the result of a quasi-linear diffusion in the velocity space, together
with the development of persistent magnetic humps. At larger distance from threshold and
for a $\beta_\|$ that is not too large, the flattening of the distribution function is
not observed and, after a while, the early formed magnetic humps can evolve toward magnetic
holes. A similar change of shape is obtained when retaining the expansion of the plasma,
which makes the system to progressively evolve to a stable regime.  Below threshold,
magnetic humps are damped but magnetic holes turn out to be persistent. This suggests
that the magnetic holes displaying the polarization properties of the mirror modes,
result from the evolution of magnetic dips and humps possibly convected to the point
of observation.

A fully systematic theory reproducing all the observational and numerical results is still
to be developed. A reductive perturbative expansion leads to an asymptotic equation
that can be viewed as an extension of the growth rate equation, retaining the
leading nonlinearity with a coupling coefficient whose sign prescribes the shape
(depressions or enhancements) of the mirror structures. In the case of a bi-Maxwellian initial
equilibrium, this approach predicts magnetic holes. It turns out that taking into
account the early quasi-linear evolution changes the sign of the nonlinear coupling
coefficient, and
thus reproduces the formation of the magnetic humps observed in the direct numerical
simulations. The reductive perturbative expansion also predicts that the bifurcation
associated with the mirror instability is subcritical, which provides an interpretation
of both the phenomenon of bistability (i.e., the stability of mirror structures below threshold)
and the large amplitude of the observed mirror structures. The latter property is
associated with the development of a finite-time singularity for asymptotic solutions
above threshold, indicating a breakdown of the asymptotic scaling. Nonlinear Landau damping
provides a  plausible mechanism to saturate the structure amplitude.

In addition to their fundamental interest, linear and nonlinear mirror modes are expected
to affect the transport properties of the plasma and also to constrain the temperature
anisotropy associated with a dominant perpendicular heating of the ions, observed in the solar wind.
As recently suggested by simulations of a refined Landau fluid model, such a heating could
result in particular from the quasi-transverse dynamics of a turbulent plasma driven by large-scale
quasi-incompressible oscillations, a scenario that is consistent with
the classical picture of a quasi-transverse Alfv\'enic cascade and with previous
suggestions of a non resonant ion heating by Alfv\'en wave turbulence.

\begin{theacknowledgments}
It is my pleasure to thank
D.~Borgogno, F.~Califano, V.~G\'enot, P.~Hellinger, E.~Kuznetsov, D.~Laveder,
L.~Marradi, T.~Passot and  V.~Ruban for a close collaboration on the issues
discussed in this paper.
Partial support by Programme National Soleil Terre of INSU-CNRS is acknowledged.
\end{theacknowledgments}

\bibliographystyle{aipproc}
\bibliography{biblio_sulem-2}

\begin{thebibliography}{112}
\expandafter\ifx\csname natexlab\endcsname\relax\def\natexlab#1{#1}\fi
\providecommand{\enquote}[1]{``#1''}
\expandafter\ifx\csname url\endcsname\relax
  \def\url#1{\texttt{#1}}\fi
\expandafter\ifx\csname urlprefix\endcsname\relax\def\urlprefix{URL }\fi
\providecommand{\eprint}[2][]{\url{#2}}

\bibitem[Soucek et~al.(2007)]{SLD07}
J.~Soucek, E.~Lucek, and I.~Dandouras, \emph{J. Geophys. Res.} \textbf{113},
  A04203 (2007).

\bibitem[Joy et~al.(2006)]{Joy06}
S.~P. Joy, M.~G. Kivelson, R.~J. Walker, K.~K. Khurana, C.~T. Russell, and
  W.~R. Paterson, \emph{J. Geophys. Res.} \textbf{111}, A12212 (2006).

\bibitem[Gary(1993)]{GaryBook93}
S.~P. Gary, \emph{Theory of Space Plasma Microinstabilities}, Cambridge
  University Press, Cambridge, 1993.

\bibitem[Lucek et~al.(2005)]{Lucek05}
E.~A. Lucek, D.~Constantinescu, M.~L. Goldstein, J.~Pickett, J.~L. Pin\c{c}on,
  F.~Sahraoui, R.~A. Treumann, and S.~N. Walker, \emph{Space Sci. Rev.}
  \textbf{118}, 95--152 (2005).

\bibitem[Shoji et~al.(2009)]{Shoji09}
M.~Shoji, Y.~Omura, B.~T. Tsurutani, O.~P. Verkhoglyadova, and B.~Lembege,
  \emph{J. Geophys. Res.} \textbf{114}, A10203 (2009).

\bibitem[Bourouaine et~al.(2010)]{Bourou10}
S.~Bourouaine, E.~Marsch, and F.~M. Neubauer, \emph{Geophys. Res. Lett.}
  \textbf{37}, L14104 (2010).

\bibitem[Stasiewicz(2004{\natexlab{a}})]{Sta04}
K.~Stasiewicz, \emph{Geophys. Res. Lett.} \textbf{31}, L21804
  (2004{\natexlab{a}}).

\bibitem[Baumgartel(2003)]{Baum03}
K.~Baumgartel, \emph{Geophys. Res. Lett.} \textbf{30}, 1761 (2003).

\bibitem[Winterhalter et~al.(1994)]{Winter94}
D.~Winterhalter, M.~Neugebauer, B.~E. Goldstein, E.~J. Smith, S.~J. Bame, and
  A.~Balogh, \emph{J. Geophys. Res.} \textbf{99}, 23371--23381 (1994).

\bibitem[Vedenov and Sagdeev(1959)]{VS59}
A.~A. Vedenov, and R.~Z. Sagdeev, \enquote{Some properties of a plasma with an
  anisotropic ion velocity distribution in a magnetic field,} in \emph{Plasma
  Physics and the Problem of Controlled Thermonuclear Reactions}, English
  translation, Pergamon, New York, 1959, vol.~3, pp. 332--339.

\bibitem[Chandrasekhar et~al.(1958)]{Chandra58}
S.~Chandrasekhar, A.~N. Kaufman, and K.~M. Watson, \emph{Proc. R. Soc. London}
  \textbf{A245}, 435--455 (1958).

\bibitem[Rudakov and Sagdeev(1959)]{RS59}
L.~I. Rudakov, and R.~Z. Sagdeev, \enquote{A quasi-hydrodynamic description of
  a rarefied plasma in a magnetic field,} in \emph{Plasma Physics and the
  Problem of Controlled Thermonuclear Reactions}, English traduction, Pergamon,
  New York, 1959, vol.~3, pp. 321--331.

\bibitem[Treumann and Baumjohan(1997)]{Treu97}
R.~A. Treumann, and W.~Baumjohan, \emph{Advanced Space Plasma Physics},
  Imperial College Press, London, 1997.

\bibitem[Southwood and Kivelson(1993)]{SK93}
D.~J. Southwood, and M.~G. Kivelson, \emph{J. Geophys. Res.} \textbf{98},
  9181--9187 (1993).

\bibitem[Hasegawa(1969)]{Hase69}
A.~Hasegawa, \emph{Phys. Fluids} \textbf{12}, 2642--2650 (1969).

\bibitem[Pokhotelov et~al.(1985)]{Pokho85}
O.~A. Pokhotelov, V.~A. Pilipenko, and E.~Amata, \emph{Planet Space Sc.}
  \textbf{33}, 1229--1241 (1985).

\bibitem[Pokhotelov et~al.(2001)]{Pokho01}
O.~A. Pokhotelov, M.~A. Balikhin, R.~A. Treumann, and V.~P. Pavlenko, \emph{J.
  Geophys. Res.} \textbf{106}, 8455--8463 (2001).

\bibitem[Shapiro and Shevchenko(1964)]{shsh64}
V.~D. Shapiro, and V.~I. Shevchenko, \emph{Sov. Phys. JETP} \textbf{18},
  1109--1116 (1964).

\bibitem[Hall(1979)]{Hall79}
A.~N. Hall, \emph{J. Plasma Phys.} \textbf{21}, 431--443 (1979).

\bibitem[Pokhotelov et~al.(2004)]{Pokho04}
O.~A. Pokhotelov, R.~Z. Sagdeev, M.~A. Balikhin, and R.~A. Treumann, \emph{J.
  Geophys. Res.} \textbf{109}, A09213 (2004).

\bibitem[Pantellini and Schwartz(1995)]{PS95}
F.~G.~E. Pantellini, and S.~J. Schwartz, \emph{J. Geophys. Res.} \textbf{100},
  3539--3549 (1995).

\bibitem[Pokhotelov et~al.(2000)]{Pokho00}
O.~A. Pokhotelov, M.~A. Balikhin, H.~S.-C.~K. Alleyne, and O.~G. Onishchenko,
  \emph{J. Geophys. Res.} \textbf{105}, 2393--2401 (2000).

\bibitem[Hellinger(2007)]{Hell07}
P.~Hellinger, \emph{Phys. Plasmas} \textbf{14}, 082105 (2007).

\bibitem[Stix(1962)]{Stix62}
T.~H. Stix, \emph{The Theory of Plasma Waves}, McGraw-Hill, New York, 1962.

\bibitem[Gary(1992)]{Gary92}
S.~P. Gary, \emph{J. Geophys. Res.} \textbf{97}, 8519--8529 (1992).

\bibitem[Passot and Sulem(2007)]{PS07}
T.~Passot, and P.~L. Sulem, \emph{Phys. Plasmas} \textbf{14}, 082502 (2007).

\bibitem[Gedalin et~al.(2001)]{Geda01}
M.~Gedalin, Y.~E. Lyubarsky, M.~Balikhin, and C.~T. Russell, \emph{Phys.
  Plasmas} \textbf{8}, 2934--2945 (2001).

\bibitem[Pokhotelov et~al.(2002)]{Pokho02}
O.~A. Pokhotelov, R.~A. Treumann, R.~Z. Sagdeev, M.~A. Balikhin, O.~G.
  Onishchenko, V.~P. Pavlenko, and I.~Sandberg, \emph{J. Geophys. Res.}
  \textbf{107}, A101312 (2002).

\bibitem[Pokhotelov et~al.(2005)]{PBST05}
O.~A. Pokhotelov, M.~A. Balikhin, R.~Z. Sagdeev, and R.~A. Treumann, \emph{J.
  Geophys. Res.} \textbf{110}, A10206 (2005).

\bibitem[Qu et~al.(2007)]{QLC07}
H.~Qu, Z.~Lin, and L.~Chen, \emph{Phys. Plasmas} \textbf{14}, 042108 (2007).

\bibitem[Kauffman et~al.(1970)]{Kauf70}
R.~L. Kauffman, J.-T. Horng, and A.~Wolfe, \emph{J. Geophys. Res., (Space
  Phys.)} \textbf{75}, 4666--4676 (1970).

\bibitem[Tsurutani et~al.(1982)]{Tsuru82}
B.~T. Tsurutani, E.~J. Smith, R.~R. Anderson, K.~W. Ogilvie, J.~D. Scudder,
  D.~N. Baker, and S.~J. Bame, \emph{J. Geophys. Res.} \textbf{87}, 6060--6072
  (1982).

\bibitem[Lacombe and Belmont(1995)]{LB95}
C.~Lacombe, and G.~Belmont, \emph{Adv. Space Res.} \textbf{15}, 329--340
  (1995).

\bibitem[Lucek et~al.(1999)]{Lucek99}
E.~A. Lucek, M.~W. Dunlop, A.~Balogh, P.~Cargill, a.~E.~G. W.~Baumjohann,
  G.~Haerendel, and K.~H. Fornacon, \emph{Geophys. Res. Lett.} \textbf{26},
  2159--2162 (1999).

\bibitem[Lucek et~al.(2001)]{Lucek01}
E.~A. Lucek, M.~W. Dunlop, T.~S. Horbury, A.~Balogh, P.~Cargill, C.~Carr, K.-H.
  Fornacon, E.~Georgescu, and T.~Oddy, \emph{Ann.Geophys.} \textbf{19},
  1421--1428 (2001).

\bibitem[T\'atrallyay and Erdos(2005)]{TE05}
M.~T\'atrallyay, and G.~Erdos, \emph{Planet Space Sc.} \textbf{53}, 33--40
  (2005).

\bibitem[G\'enot et~al.(2009)]{Genot09}
V.~G\'enot, E.~Budnik, P.~Hellinger, T.~Passot, G.~Belmont, P.~M.
  Tr\'evn\'i\u{c}ek, P.~L. Sulem, E.~Lucek, and I.~Dandouras, \emph{Ann.
  Geophys} \textbf{27}, 601--615 (2009).

\bibitem[Horbury and Lucek(2009)]{HL09}
T.~S. Horbury, and E.~A. Lucek, \emph{J. Geophys. Res.} \textbf{114}, A052217
  (2009).

\bibitem[Rae et~al.(2007)]{Rae07}
I.~J. Rae, I.~R. Mann, C.~E.~J. Watt, L.~M. Kistler, and W.~Baumjohann,
  \emph{J. Geophys. Res.} \textbf{112}, A11203 (2007).

\bibitem[Erd\"os and Balogh(1996)]{EB96}
G.~Erd\"os, and A.~Balogh, \emph{J. Geophys. Res.} \textbf{101}, 1--12 (1996).

\bibitem[Andr\'e et~al.(2002)]{Andre02}
N.~Andr\'e, G.~Erd\"os, and M.~Dougherty, \emph{Geophys. Res. Lett.}
  \textbf{29}, 1980 (2002).

\bibitem[Violante et~al.(1995)]{Violante95}
L.~Violante, M.~B. Bavassano-Cattaneo, G.~Moreno, and J.~D. Richardson,
  \emph{J. Geophys. Res.} \textbf{100}, 12047--12055 (1995).

\bibitem[Bavassano-Cattaneo et~al.(1998)]{Bavas98}
M.~B. Bavassano-Cattaneo, C.~Basile, G.~Moreno, and J.~D. Richardson, \emph{J.
  Geophys. Res.} \textbf{103}, 11961--11972 (1998).

\bibitem[Volwerk et~al.(2008{\natexlab{a}})]{Volwerk08a}
M.~Volwerk, T.~L. Zhang, M.~Delva, Z.~V\"or\"os, W.~Baumjohann, and K.-H.
  Glassmeier, \emph{Geophys. Res. Lett.} \textbf{35}, L12204
  (2008{\natexlab{a}}).

\bibitem[Volwerk et~al.(2008{\natexlab{b}})]{Volwerk08b}
M.~Volwerk, T.~L. Zhang, M.~Delva, Z.~V\"or\"os, W.~Baumjohann, and K.-H.
  Glassmeier, \emph{J. Geophys. Res.} \textbf{113}, E00B16
  (2008{\natexlab{b}}).

\bibitem[Russell et~al.(1999)]{Russell99}
C.~T. Russell, E.~Huddleston, R.~J. Strangeway, X.~Blanco-Cano, M.~G. Kivelson,
  K.~K. Khurana, L.~A. Frank, W.~Paterson, D.~A. Gurnett, and W.~S. Kurth,
  \emph{J. Geophys. Res.} \textbf{104}, 17471--17477 (1999).

\bibitem[Huddleston et~al.(1999)]{Hudd99}
E.~Huddleston, R.~J. Strangeway, X.~Blanco-Cano, C.~T. Russell, M.~G. Kivelson,
  and K.~K. Khurana, \emph{J. Geophys. Res.} \textbf{104}, 17479--17489 (1999).

\bibitem[Russell(1987)]{Russell87}
C.~T. Russell, \emph{Geophys. Res. Lett.} \textbf{14}, 644--647 (1987).

\bibitem[Tsurutani et~al.(1999{\natexlab{a}})]{Tsuru99a}
B.~T. Tsurutani, G.~S. Lakhina, E.~J. Smith, B.~Buti, S.~L. Moses, F.~V.
  Coroniti, A.~L. Brinca, J.~A. Slavin, and R.~D. Zwickl, \emph{Nonlin. Proc.
  Geophys.} \textbf{6}, 229--234 (1999{\natexlab{a}}).

\bibitem[Zhang et~al.(2008)]{Zhang08}
T.~L. Zhang, C.~T. Russell, W.~Baumjohann, L.~K. Jian, M.~A. Balikhin, J.~B.
  Cao, C.~Wang, X.~Blanco-Cano, K.-H. Glassmeier, W.~Zambelli, M.~Volwerk,
  M.~Delva, and Z.~V\"or\"os, \emph{Geophys. Res. Lett.} \textbf{35}, L10106
  (2008).

\bibitem[Zhang et~al.(2009)]{Zhang09}
T.~L. Zhang, W.~Baumjohann, C.~T. Russell, L.~K. Jian, C.~Wang, J.~B. Cao,
  M.~A. Balikhin, X.~Blanco-Cano, M.~Delva, and Z.~Volwerk, \emph{J. Geophys.
  Res.} \textbf{114}, A10107 (2009).

\bibitem[Burlaga et~al.(2006)]{Burl06}
L.~F. Burlaga, N.~F. Ness, and M.~H. Acuna, \emph{Geophys. Res. Lett.}
  \textbf{33}, L21106 (2006).

\bibitem[Burlaga et~al.(2007)]{Burl07}
L.~F. Burlaga, N.~F. Ness, and M.~H. Acuna, \emph{J. Geophys. Res.}
  \textbf{112}, A07106 (2007).

\bibitem[Tsurutani et~al.(1992)]{Tsuru92}
B.~Tsurutani, D.~J. Southwood, E.~J. Smith, and A.~Balogh, \emph{Geophys. Res.
  Lett.} \textbf{19}, 1267--1270 (1992).

\bibitem[Liu et~al.(2006)]{Liu06}
Y.~Liu, J.~D. Richardson, J.~W. Belcher, J.~C. Kasper, and R.~M. Skoug,
  \emph{J. Geophys. Res.} \textbf{111}, A09108 (2006).

\bibitem[Tsurutani et~al.(1999{\natexlab{b}})]{Tsuru99b}
B.~T. Tsurutani, G.~S. Lakhina, D.~Winterhalter, J.~K. Arballo, C.~Galvan, and
  R.~Sukurai, \emph{Nonlin. Proc. Geophys.} \textbf{6}, 235--242
  (1999{\natexlab{b}}).

\bibitem[Schekochihin et~al.(2008)]{Scheko08}
A.~A. Schekochihin, S.~C. Cowley, R.~M. Kulsrud, M.~S. Rosin, and T.~Heinemann,
  \emph{Phys. Rev. Lett.} \textbf{100}, 081301 (2008).

\bibitem[Horbury et~al.(2004)]{HLBDR}
T.~S. Horbury, E.~A. Lucek, A.~Bulogh, I.~Dandouras, and H.~R\`eme, \emph{J.
  Geophys. Res.} \textbf{109}, A09202 (2004).

\bibitem[G\'enot et~al.(2001)]{Genot01}
V.~G\'enot, S.~J. Schwartz, C.~Mazelle, M.~Bakhilin, M.~Dunlop, and T.~M.
  Bauer, \emph{J. Geophys. Res.} \textbf{106}, 21611--21622 (2001).

\bibitem[G\'enot et~al.(2007)]{Genot07}
V.~G\'enot, E.~Budnik, C.~Jacquey, I.~Dandouras, and E.~Lucek, \enquote{Mirror
  modes observed with Cluster in the earth's magnetosheath: statisitcal study
  and IMF/solar wind dependence,} in \emph{Advances in Geosciences}, edited by
  M.~Duldig, World Scientific Publishing Company, Singapore, 2007, vol.~14, pp.
  263--283.

\bibitem[Leckband et~al.(1995)]{Leck95}
J.~A. Leckband, D.~Burgess, F.~G.~E. Pantellini, and S.~J. Schwartz, \emph{Adv.
  Space Res.} \textbf{15}, 345--348 (1995).

\bibitem[Phan et~al.(1994)]{Phan94}
T.~D. Phan, W.~B. G.~Paschmann, N.~Sckopke, and H.~L\"uhr, \emph{J. Geophys.
  Res.} \textbf{99}, 121--141 (1994).

\bibitem[Baumg\"artel(1999)]{Baumg99}
K.~Baumg\"artel, \emph{J. Geophys. Res.} \textbf{104}, 28295--28308 (1999).

\bibitem[Winterhalter et~al.(2000)]{Winter00}
D.~Winterhalter, E.~J. Smith, M.~Neugebauer, B.~E. Goldstein, and B.~T.
  Tsurutani, \emph{Geophys. Res. Lett.} \textbf{27}, 1615--1618 (2000).

\bibitem[Stevens and Kasper(2007)]{SK07}
M.~L. Stevens, and J.~C. Kasper, \emph{J. Geophys. Res.} \textbf{112}, A05109
  (2007).

\bibitem[Buti et~al.(2001)]{Buti01}
B.~Buti, B.~T. Tsurutani, M.~Neugebauer, and B.~E. Goldstein, \emph{Geophys.
  Res. Lett.} \textbf{28}, 1355--1358 (2001).

\bibitem[Schwartz et~al.(1996)]{Schwartz96}
S.~J. Schwartz, D.~Burgess, and J.~J. Mose, \emph{Ann. Geophys.} \textbf{14},
  1134--1150 (1996).

\bibitem[Price et~al.(1986)]{Price86}
C.~P. Price, D.~W. Swift, and L.-C. Lee, \emph{J. Geophys. Res.} \textbf{91},
  101--112 (1986).

\bibitem[Gary et~al.(1993)]{Gary93}
S.~P. Gary, S.~A. Fuselier, and B.~J. Anderson, \emph{J. Geophys. Res.}
  \textbf{98}, 1481--1488 (1993).

\bibitem[McKean et~al.(1994)]{McKean94}
M.~E. McKean, D.~Winske, and S.~P. Gary, \emph{J. Geophys. Res.} \textbf{99},
  11141--11153 (1994).

\bibitem[Anderson et~al.(1993)]{Anderson94}
B.~J. Anderson, S.~A. Fuselier, S.~P. Gary, and R.~E. Denton, \emph{J. Geophys.
  Res.} \textbf{99}, 5877--5891 (1993).

\bibitem[Czaykowska1 et~al.(2001)]{Czayko}
A.~Czaykowska1, T.~M. Bauer, R.~A. Treumann, and W.~Baumjohann, \emph{Ann.
  Geophys.} \textbf{19}, 275--287 (2001).

\bibitem[Winterhalter et~al.(1995)]{Winter95}
D.~Winterhalter, M.~Neugebauer, B.~E. Goldstein, E.~J. Smith, B.~T. Tsurutani,
  S.~J. Bame, and A.~Balogh, \emph{Space Sci. Rev.} \textbf{72}, 201--204
  (1995).

\bibitem[Sperveslage et~al.(2000)]{Sper00}
K.~Sperveslage, F.~M. Neubauer, K.~Baumg\"artel, and N.~F. Ness, \emph{Nonlin.
  Proc. Geophys.} \textbf{7}, 191--200 (2000).

\bibitem[Chew et~al.(1956)]{CGL}
C.~F. Chew, M.~L. Goldberger, and F.~E. Low, \emph{Proc. Roy. Soc. (London)}
  \textbf{A236}, 112--118 (1956).

\bibitem[Kulsrud(1983)]{Kulsrud83}
R.~M. Kulsrud, \emph{MHD description of plasma, {\rm in} Handbook of Plasma
  Physics, {\rm edited by M.~N.~Rosenbluth and R.~Z.~Sagdeev}}, North-Holland
  Publishing Company, New York, 1983, vol. 1: Basic Plasma Physics I, edited by
  A.~A.~Galeev and R.~N.~Sudan, chap. 1.4, pp. 115--145.

\bibitem[Snyder et~al.(1997)]{SHD97}
P.~B. Snyder, G.~W. Hammett, and W.~Dorland, \emph{Phys. Plasmas} \textbf{4},
  3974--3985 (1997).

\bibitem[Hau et~al.(2005)]{Hau05}
L.-N. Hau, B.-J. Wang, and W.-L. Teh, \emph{Phys. Plasmas} \textbf{12}, 122904
  (2005).

\bibitem[Hau and Wang(2007)]{Hau07}
L.-N. Hau, and B.-J. Wang, \emph{Nonlin. Proc. Geophys.} \textbf{14}, 557--568
  (2007).

\bibitem[Passot et~al.(2005)]{PRS06}
T.~Passot, V.~Ruban, and P.~L. Sulem, \emph{Phys. Plasmas} \textbf{13}, 102310
  (2005).

\bibitem[Constantinescu(2002)]{Const02}
O.~D. Constantinescu, \emph{J. Atm. Solar-Terrestrial Phys.} \textbf{64},
  645--649 (2002).

\bibitem[Califano et~al.(2008)]{Calif08}
C.~Califano, P.~Hellinger, E.~Kuznetsov, T.~Passot, P.~L. Sulem, and P.~M.
  {Tr\'avn\'{\i}\v{c}ek}, \emph{J. Geophys. Res.} \textbf{113}, A08219 (2008).

\bibitem[{Tr\'avn\'{\i}\v{c}ek} et~al.(2007)]{Trav07}
P.~M. {Tr\'avn\'{\i}\v{c}ek}, P.~Hellinger, M.~G.~G.~T. Taylor, C.~P. Escoubet,
  I.~Dandouras, and E.~Lucek, \emph{Geophys. Res. Lett.} \textbf{34}, L15104
  (2007).

\bibitem[Kuznetsov et~al.(2007{\natexlab{a}})]{KPS07a}
E.~A. Kuznetsov, T.~Passot, and P.-L. Sulem, \emph{Phys. Rev. Lett.}
  \textbf{98}, 235003 (2007{\natexlab{a}}).

\bibitem[Hellinger et~al.(2009)]{HKP09}
P.~Hellinger, E.~Kuznetsov, T.~Passot, P.~L. Sulem, and P.~Tr{\'a}vn{\'\i}{\v
  c}ek, \emph{Geophys. Res. Lett.} \textbf{36}, L06103 (2009).

\bibitem[Pokhotelov et~al.(2008)]{Pokho08}
O.~A. Pokhotelov, R.~Z. Sagdeev, M.~A. Balikhin, O.~G. Onishchenko, and V.~N.
  Fedun, \emph{J. Geophys. Res.} \textbf{113}, A04225 (2008).

\bibitem[Kuznetsov et~al.(2007{\natexlab{b}})]{KPS07b}
E.~A. Kuznetsov, T.~Passot, and P.-L. Sulem, \emph{JETP Letters} \textbf{86},
  637--642 (2007{\natexlab{b}}).

\bibitem[Pantellini et~al.(1995)]{pant95}
F.~G.~E. Pantellini, D.~Burgess, and S.~J. Schwartz, \emph{Adv. Space Res.}
  \textbf{15}, 341--344 (1995).

\bibitem[Kivelson and Southwood(1996)]{kiso96}
M.~G. Kivelson, and D.~J. Southwood, \emph{J. Geophys. Res.} \textbf{101},
  17365--17372 (1996).

\bibitem[Pantellini(1998)]{pant98}
F.~G.~E. Pantellini, \emph{J. Geophys. Res.} \textbf{103}, 4789--4798 (1998).

\bibitem[Diamond et~al.(2010)]{DII10}
P.~H. Diamond, S.-I. Itoh, and K.~Itoh, \emph{Modern Plasma Physics; Volume1:
  Physical Kinetics of Turbulent Plasmas}, Cambridge University Press,
  Cambridge, 2010.

\bibitem[Passot et~al.(2009)]{PSKH09}
T.~Passot, P.~L. Sulem, E.~Kuznetsov, and P.~Hellinger, \enquote{Influence of
  kinetic effects on the shape of mirror structures,} in \emph{New developments
  in Nonlinear Plasma Physics, (Proceedings of the 2009 ICTP Summer College on
  Plasma Physics and International Symposium on Cutting Edge Plasma Physics,
  Trieste, Italy, 24--28 August 2009)}, edited by B.~Eliasson, and P.~K.
  Shukla, AIP Conference Proceedings 1188, American Institute of Physics, 2009,
  pp. 205--212.

\bibitem[Istomin et~al.(2009)]{IPB09}
Y.~N. Istomin, O.~A. Pokhotelov, and M.~A. Balikhin, \emph{Phys. Plasmas}
  \textbf{16}, 062905 (2009).

\bibitem[Jovanovi\'c and Shukla(2009)]{JS09}
D.~Jovanovi\'c, and P.~K. Shukla, \emph{Phys. Plasmas} \textbf{16}, 082901
  (2009).

\bibitem[Sahraoui et~al.(2006)]{Sahra06}
F.~Sahraoui, G.~Belmont, L.~Rezeau, N.~Cornilleau-Wehrlin, J.~L. Pin\c{c}on,
  and A.~Balogh, \emph{Phys. Rev. Lett.} \textbf{96}, 075002 (2006).

\bibitem[Pin\c{c}on and Lefeuvre(1991)]{PL91}
J.~L. Pin\c{c}on, and F.~Lefeuvre, \emph{J. Geophys. Res.} \textbf{96},
  1789--1802 (1991).

\bibitem[Walker et~al.(2004)]{Walker04}
S.~N. Walker, F.~Sahraoui, M.~A. Balikhin1, G.~Belmont, J.~L. Pincon,
  L.~Rezeau, H.~Alleyne, N.~Cornilleau-Wehrlin, and M.~Andr\'e, \emph{Ann.
  Geophys.} \textbf{22}, 3021--3032 (2004).

\bibitem[Hellinger et~al.(2006)]{Hell06}
P.~Hellinger, P.~{Tr\'avn\'{\i}\v{c}ek}, J.~C. Kasper, and A.~J. Lazarus,
  \emph{Geophys. Res. Lett.} \textbf{33}, L09101 (2006).

\bibitem[Bale et~al.(2009)]{Bale09}
S.~D. Bale, J.~C. Kasper, G.~G. Howes, E.~Quataert, C.~Salem, and D.~Sundkvist,
  \emph{Phys. Rev. Lett.} \textbf{103}, 211101 (2009).

\bibitem[Hammett and Perkins(1990)]{HP90}
G.~W. Hammett, and F.~W. Perkins, \emph{Phys. Rev. Lett.} \textbf{64},
  3019--3022 (1990).

\bibitem[Passot and Sulem(2003)]{PS03}
T.~Passot, and P.~L. Sulem, \emph{Phys. Plasmas} \textbf{10}, 3906--3913
  (2003).

\bibitem[Goswami et~al.(2005)]{GPS05}
P.~Goswami, T.~Passot, and P.~L. Sulem, \emph{Phys. Plasmas} \textbf{12},
  102109 (2005).

\bibitem[Borgogno et~al.(2007)]{BPS07}
D.~Borgogno, T.~Passot, and P.~L. Sulem, \emph{Nonlin. Proc. Geophys.}
  \textbf{14}, 373--383 (2007).

\bibitem[Tsurutani et~al.(2002)]{Tsuru02}
B.~T. Tsurutani, B.~Dasgupta, C.~Galvan, M.~Neugebauer, G.~S. Lakhina, J.~K.
  Arballo, D.~Winterhalter, B.~E. Goldstein, and B.~Buti, \emph{Geophys. Res.
  Lett.} \textbf{29}, 2233 (2002).

\bibitem[Stasiewicz(2004{\natexlab{b}})]{Sta04a}
K.~Stasiewicz, \emph{Phys. Rev. Lett.} \textbf{93}, 125004
  (2004{\natexlab{b}}).

\bibitem[Stasiewicz(2005)]{Sta05}
K.~Stasiewicz, \emph{J. Geophys. Res.} \textbf{110}, A03220 (2005).

\bibitem[Mj{\o}lhus(2006)]{Mjolhus06}
E.~Mj{\o}lhus, \emph{Phys. Scr.} \textbf{T122}, 135--153 (2006).

\bibitem[Mj{\o}lhus(2009)]{Mjolhus09}
E.~Mj{\o}lhus, \emph{Nonlin. Proc. Geophys.} \textbf{16}, 251--264 (2009).

\bibitem[Baumg\"artel et~al.(2005)]{BSD05}
K.~Baumg\"artel, K.~Sauer, and E.~Dubinin, \emph{Nonlin. Proc. Geophys.}
  \textbf{12}, 291--298 (2005).

\bibitem[Laveder et~al.(2010)]{LMPS10}
D.~Laveder, L.~Marradi, T.~Passot, and P.~L. Sulem, Turbulence-driven
  temperature anisotropies and constraining effects of mirror instability in a
  magnetized plasma: Landau fluid simulations (2010), in preparation.

\bibitem[Bourouaine et~al.(2008)]{Bourouaine08}
S.~Bourouaine, E.~Marsch, and C.~Vocks, \emph{Astrophys. J.} \textbf{684},
  L119--L122 (2008).

\bibitem[Chandran et~al.(2010)]{Chandran10}
B.~D.~G. Chandran, B.~Li, B.~N. Rogers, E.~Quataert, and K.~Germashchewski,
  \emph{Astrophys. J.} \textbf{720}, 503--515 (2010).

\end{thebibliography}
\end{document}